\documentclass[prb,amsmath,amssymb,superscriptaddress,twocolumn]{revtex4}
\usepackage{graphicx}
\usepackage[usenames]{color}

\newcommand{\br}{{\bf r}}
\newcommand{\bx}{{\bf x}}
\newcommand{\by}{{\bf y}}
\newcommand{\bk}{{\bf k}}

\newcommand{\bq}{{\bf q}}

\newcommand{\bK}{{\bf K}}

\newcommand{\bR}{{\bf R}}

\newcommand{\ms}{{m^{\ast}}}

\newcommand{\eps}{\epsilon}

\DeclareMathAlphabet{\mathpzc}{OT1}{pzc}{m}{it}

\newcommand {\rmd}{{\rm d}}

\begin{document}
\title{Electronic multicriticality in bilayer graphene}
\author{Vladimir Cvetkovic}
\author{Robert E.\ Throckmorton}
\author{Oskar Vafek}
\affiliation{National High Magnetic Field Laboratory and Department
of Physics,\\ Florida State University, Tallahasse, Florida 32306,
USA}

\date{\today}
\begin{abstract}
We map out the possible ordered states in bilayer graphene at the
neutrality point by extending the previous renormalization group
treatment of many-body instabilities to finite temperature, trigonal
warping and externally applied perpendicular electric field. We were
able to analytically determine all outcomes of the RG flow equations
for the nine four-fermion coupling constants. While the full phase
diagram exhibits a rich structure, we confirm that when forward
scattering dominates, the only ordering tendency with divergent
susceptibility at finite temperature is the nematic. At finite
temperature, this result is stable with respect to small back and
layer imbalance scattering; further increasing their strength leads
to the layer antiferromagnet. We also determine conditions for other
ordered states to appear and compare our results to the special
cases of attractive and repulsive Hubbard models where exact results
are available.
\end{abstract}

\maketitle

\section{Introduction}

Understanding itinerant electronic systems with competing ordering
tendencies is among the most profound challenges in today's
condensed matter theory. In one-dimensional systems, powerful
theoretical tools are available for answering some of the
questions\cite{GiamarchiBook}, but extending the techniques to
higher dimensions has met with limited success. Often the problem is how
to treat the various ordering tendencies on equal footing without an
inherent bias towards any one of the possible ordered states.

In this regard, bilayer graphene at, and near, the neutrality point
can be regarded as a model system. To a first approximation,
there is a conduction band and a valence band that touch quadratically
near two points, $\bK$ and $\bK'=-\bK$, in the Brillouin zone \cite
{McCannFalkoPRB2006,CastroNetoRMP2009}. Even when all
electron-electron interactions are ignored, such a system would have
low-temperature susceptibilities which diverge as $\sim\ln T$
towards a number of different ordered states. While there are no
known exact solutions, such a situation is expected to lead to
instabilities with respect to infinitesimal electron-electron
interactions. The challenge is then to identify the conditions under
which any one combination of the various possible states gets
preferably selected as the temperature is lowered.

Since a many-body ordering appears already at weak coupling this
problem is amenable to the renormalizaton group (RG) approach, whose
advantage is that it can account for the competing tendencies in an
unbiased way. Moreover, since the few-milli-electron-volt energy scales associated
with ordering extracted from present-day
experiments\cite{YacobyScience2010, YacobyPRL2010, Mayorov2011,VelascoNatNano2012,
FreitagPRL2012,VeliguraPRB2012,BaoArXiv}
are much smaller than the natural upper cutoff in the problem
originating from the split-off bands derived from the dimerized sites
($\sim200-300$ meV), the physical system itself is expected to be
well described by a weak coupling theory. Therefore, we expect that
the competition among the number of inherently strong-coupling
phases can be accessed within such a weak coupling approximation.

The previous RG treatments of this problem presented in Refs.\
\onlinecite{VafekYangPRB2010,VafekPRB2010} consisted of first
building a low-energy effective field theory, which, when
electron-electron interactions are neglected, can be thought of as a
Gaussian fixed point of the RG scale transformation \cite
{ShankarRMP1994} with dynamical critical exponent $z=2$. Except
under some non-generic fine-tuned initial conditions, contact
interactions have been shown to be marginally relevant at this fixed
point. Such four-fermion terms in the low-energy effective field
theory arise from microscopic electron-electron interactions
$V_{ee}(\br)$ whose Fourier transform is non-divergent in the small
wavevector limit. They could, for instance, correspond to $1/r$
Coulomb interactions screened by proximity to metallic gates. Within
this approach, the electronic modes with momenta in a thin shell
$(1-\Delta\ell)\Lambda<|\bk|<\Lambda$ near the cutoff $\Lambda$, and
arbitrary frequency $\omega$, are integrated out, while the change
in the effective action is monitored as the process is
iterated\cite{WilsonPRB1971,ShankarRMP1994}. To determine the
leading instability, infinitesimal symmetry breaking source terms
were introduced \cite
{VafekYangPRB2010,VafekPRB2010,ThrockmortonArXiv} and included in
the process of renormalization. The source term with the strongest
divergence was then identified as the most dominant ordering
tendency. In the case of purely forward scattering, or, in the
notation of this paper, for $g_{A_{1g}}$ only, the leading
instability was found to be toward the electronic nematic state.
This state is gapless, with either two or four Dirac points near
each $\bK$-point depending on the strength of the order parameter.
In the case of the Hubbard model, there is additional back
scattering, $g_{E_{\bK}}=\frac{1}{2}g_{A_{1g}}$, and layer imbalance
scattering $g_{A_{2u}}=g_{A_{1g}}$, and the leading instability is
found to be toward the layer antiferromagnetic state. The
single-particle (electronic) spectrum of this phase is gapped.

In a similar approach\cite{LemonikPRB2010,LemonikArXiv}, the $1/r$
Coulomb interactions among the electrons in bilayer graphene were
first screened using RPA, and then the full $\bq$- and $\omega$-dependent
effective interaction was used as the initial
condition for the subsequent Wilson-like RG treatment. While
said approach can be criticized on the grounds that the screening of the long-range
tail of the Coulomb interaction originates from integrating
out electrons all the way down to the Fermi energy, which are double counted when
reintroduced for the RG treatment, the results obtained using this
approach are in qualitative agreement with the results obtained
previously\cite{VafekYangPRB2010,VafekPRB2010}.

To this end, we present an extension of the previous RG treatment of the problem to finite
temperature\cite{ChakravartyPRB1988, MillisPRB1993,VafekAPS2012} and
finite externally applied perpendicular electric
field. This allows us to include the competition between
broken-symmetry phases with gaps in the electronic spectrum, which
may be energetically favorable, and gapless states, which may be
entropically favorable. We also study the gradual suppression of an
ordered state as the externally applied electric field is increased.
Since temperature is treated explicitly, we can obtain the
transition temperature directly, without making any of the {\it ad hoc}
assumptions inherent in translating the value of the RG scale $\ell$ at
which the couplings diverge into temperature.

As has been noted early on in the context of one-dimensional
electron systems\cite{BychkovGorkovDzyaloshinskii1966,DzyaloshinskiiLarkin1972}, it is very useful
to compare the results of an approximate RG approach to known exact
results\cite{Gaudin1967,CNYang1967,LiebWu}. Despite the
scarcity of exact results in higher dimensions, we can compare our
results to some of the non-trivial properties of the Hubbard model
at half-filling, which can be either established
exactly\cite{CNYangPRL1989,SCZhangPRL1990,CNYangSCZhangMPLB1990,AuerbachBook} or
can be obtained from Monte Carlo simulations \cite {HirschPRL1989, MoreoPRB1990, MoreoPRL1991}. In this regard, it was
shown in Ref.\ \onlinecite{VafekPRB2010} that starting with a repulsive
Hubbard model on a honeycomb bilayer lattice at half-filling for $U\ll t_{\perp}\lesssim t$ leads to the layer
antiferromagnet as the most dominant instability. In this work, we
confirm the previous finding using the finite-temperature RG scheme.
We further establish that, if we fix the value of the nine four-fermion
coupling constants to correspond to the values derivable from the Hubbard
model, then the low-energy effective field theory possesses the
SO(4) symmetry of the microscopic Hubbard
model\cite{CNYangSCZhangMPLB1990}. As a consequence, for an attractive
Hubbard model the result of our (approximate) RG analysis recovers
the {\it exact} result that the $s$-wave
superconducting order parameter can be continuously rotated to the ``CDW'' order
parameter\cite{CNYangPRL1989,SCZhangPRL1990}. Since, for bilayer
graphene, the charge-ordered state does not break the discrete
translational symmetry of the lattice, it is not strictly a density wave, but
rather corresponds to the layer-polarized state (LP). This can be
seen in our RG equations; the LP and $s_{++}$ superconducting
source terms are identical provided that we start with the values
of the four-fermion coupling constants corresponding to the Hubbard
model with $U\ll t_{\perp}\lesssim t$. Moreover, since, in the weak coupling limit, we can map the
microscopic lattice interactions to the four-fermion coupling constants in the
continuum effective field theory, we can ask what happens when we
add a $b_1$-$b_2$ interaction $V$ in addition to the on-site attraction
$U$ (see Fig.\ \ref {BLGLatticeDiagram}). When $V$ is repulsive (attractive) we find that the exact
degeneracy between the LP and $s_{++}$ SC states is lifted in favor of the LP
($s_{++}$ SC) as expected\cite{VarmaPRL1988,AuerbachBook}.

Among the differences between our present approach and the related
weak coupling approach employed in Refs.\
\onlinecite{LemonikPRB2010,LemonikArXiv} is the fact that we perform our
analysis at finite temperature, which leads to different RG equations
for the couplings than at zero temperature. In addition to the
advantages mentioned above, this allows us to systematically
determine all possible outcomes of the RG equations in the
nine-dimensional space of initial couplings. We also avoid screening
the Coulomb interaction with the bilayer graphene low-energy degrees of freedom that
enter our Wilson RG analysis. Rather, we assume that it is screened due to either
finite temperature or the presence of external metallic
gates. Finally, we do not rely on mean-field theory to determine
the phases either directly from the bare couplings (i.e., without RG)
\cite {MinPRB2008,ZhangPRB2010,NandkishorePRL2010,NandkishorePRB2010,JungPRB2011,GorbarArXiv}, or on a renormalized mean field
treatment\cite{LemonikArXiv}.  The shortcomings of other approaches\cite{NandkishorePRL2010,MinPRB2008}
have been discussed in Ref. \onlinecite{LemonikArXiv}.

Within this formulation, as shown later in the text, the flow
equations for the nine\cite{LemonikPRB2010,VafekPRB2010} coupling
constants contain additional thermal factors, with an effective
temperature $T$ that grows under RG as $e^{2\ell}$.
The flow equations \eqref{eq:gflow} for the coupling constants
describe two competing tendencies---the term proportional to a product of
two four-fermion coupling constants tends to enhance their growth, while the
thermal factors suppress the flow of the coupling. For any fixed
initial couplings and at a high enough temperature, the couplings
saturate to finite values as $\ell\rightarrow \infty$. As the
temperature is lowered, the coupling constants saturate at higher,
but still finite, values. At the transition temperature, $T_c$, the
coupling constants diverge as $\ell\rightarrow \infty$. Below $T_c$,
the coupling constants diverge at a finite value of $\ell$. The effects of
trigonal warping, parametrized by a velocity $v_3$, can be readily
included within this formalism as well \cite {LemonikPRB2010, LemonikArXiv}.
Like temperature,
trigonal warping tends to suppress the flow of the couplings. As a
result, even at $T=0$, a critical coupling strength must be exceeded
for a phase transition to occur\cite{VafekYangPRB2010,LemonikPRB2010}.
The strength of the critical coupling vanishes as $v_3$ vanishes.

The RG flow equations of the (infinitesimal) source terms for a
multitude of symmetry-breaking order parameters reveal that, at
the transition temperature, the source terms $\Delta$ acquire an
anomalous dimension, $\eta_{\Delta}$. Analysis of the free
energy correction to $\mathcal{O}\left(\Delta^2\right)$ further
reveals that, within this approximation, the physical susceptibility
for a particular $\Delta$ diverges as $T\rightarrow T_c$ if
$\eta_{\Delta}>1$. Using this condition, we determine the phase
diagram for different initial couplings (see Fig.\ \ref {FigPhaseDiagramTv3}).

We find that, for purely forward electron-electron scattering,
$g_{A_{1g}}$, the only order parameter with a divergent
susceptibility at finite $T$ is the nematic. Moreover, this is
stable with respect to the presence of small, but finite, back
scattering $g_{E_{\bK}}$ and layer imbalance scattering (i.e., the
difference between intra- and interlayer scattering) $g_{A_{2u}}$.
Performing the analysis at finite temperature is crucial for
revealing this stability. Upon increasing the back and layer
imbalance scattering, the only other divergent susceptibility is
toward a layer antiferromagnetic (AF) state. Reversing the sign of
the back scattering while fixing the layer imbalance scattering
results in a quantum spin Hall state (QSH). Reversing the sign of
the layer imbalance scattering while fixing the sign of the back
scattering gives us a layer-polarized state (LP). Reversing the sign
of both may lead to an $s$-wave superconductor. For small
$g_{A_{2u}}/g_{E_\bK}$ and $g_{E_\bK} \approx g_{A_{1g}}
>0$, we may find a Kekul\'e current state (KC). These results are
summarized in Fig.\ \ref {FigPhaseDiagramTv3}, which shows the phase
diagram in the space of initial $g_{A_{1g}}$, $g_{E_{\bK}}$, and
$g_{A_{2u}}$.

Remarkably, the flow equations for the nine coupling constants can be
analyzed in their entirety at $T_c$. We find that, if a coupling
constant diverges, it grows as $e^{2\ell}$. At the same time, the
ratios of the coupling constants may either approach values
determined by a two-parameter family of functions, which we call the
target plane, or four isolated fixed ratios that do not belong
to the fixed plane. For each of these cases we determine the
symmetry-breaking channels with divergent susceptibilities at $T_c$.
The results are summarized in Fig.\ \ref{TargetPlane_PD}.

The rest of the paper is organized as follows.  In Sec.\ II, we
present our model for the system.  Section III is dedicated to the
thorough analysis of the RG equations and our main results.
Section IV deals with the effects of an applied
perpendicular electric field on the phase boundaries.  Our conclusions
are presented in Sec.\ V.  We give details of our derivations in the
appendixes.

\section{Hamiltonian}
\begin{figure}[ht]
\centering
\includegraphics[width=\columnwidth]{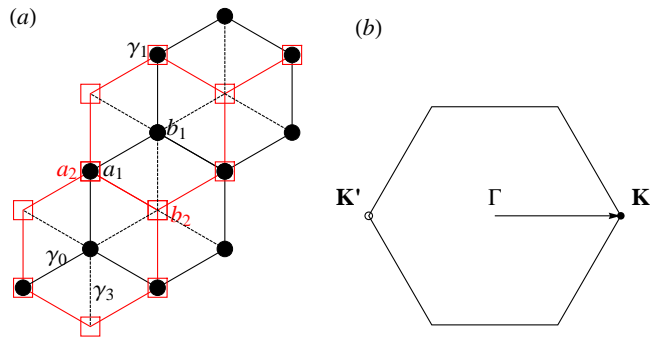}
\caption{\label{BLGLatticeDiagram}(a) The honeycomb bilayer lattice formed by
bilayer graphene.  We represent the bottom layer, 1, with black circles and the
top layer, 2, with red squares.  The $a_i$ sites are the dimerized sites, and
the $b_i$ sites are the non-dimerized sites.  We include the nearest-neighbor
intralayer hopping $\gamma_0$, the hopping between dimerized sites $\gamma_1$,
and the nearest-neighbor interlayer hopping between non-dimerized sites $\gamma_3$.
(b) The Brillouin zone associated with the honeycomb bilayer with the parabolic
degeneracy points $\bK=\frac{4\pi}{3\sqrt{3}a}\hat{\bx}$ and $\bK'=-\bK$ marked.}
\end{figure}

We will be employing a low-energy effective theory for the bilayer graphene
lattice.  This lattice and the associated Brillouin zone are shown in Fig.\
\ref{BLGLatticeDiagram}.  Our model includes the nearest-neighbor intralayer
hopping $\gamma_0\equiv t$, the hopping between dimerized sites $\gamma_1\equiv t_{\perp}$, and the
nearest-neighbor interlayer hopping between non-dimerized sites $\gamma_3$.
It is this last hopping that is responsible for trigonal warping.
Experimentally\cite{ZhangPRB2008}, $\gamma_0\approx 3\text{ eV}$, $\gamma_1\approx 0.4\text{ eV}$,
and $\gamma_3\approx 0.3\text{ eV}$.  Throughout this paper, we will use units
in which $k_B=\hbar=1$.
\subsection{Non-interacting Hamiltonian}
The tight-binding model for the lattice described above is\cite{CastroNetoRMP2009}
\begin{eqnarray}
H_{tb}=&-&\gamma_0\sum_{\bR,\delta,\sigma}(a^\dag_{1\sigma}(\bR)b_{1\sigma}(\bR+\delta)+a^\dag_{2\sigma}(\bR)b_{2\sigma}(\bR-\delta) \cr
&+&\text{h.c.}) \cr
&-&\gamma_1\sum_{\bR,\sigma}(a^\dag_{1\sigma}(\bR)a_{2\sigma}(\bR)+\text{h.c.}) \cr
&-&\gamma_3\sum_{\bR,\delta',\sigma}(b^\dag_{1\sigma}(\bR+\delta)b_{2\sigma}(\bR+\delta+\delta')+\text{h.c.}), \label {tbHamiltonian}
\end{eqnarray}
where $(a,b)_{m\sigma}(\br)$ annihilates an electron on the $(a,b)$ sublattice
on layer $m$ and site $\br$ with spin $\sigma$.  The vectors $\bR$ are the positions
of the dimerized sites within a unit cell, and $\delta$ represents one of three
vectors connecting an $a_1$ site with a nearest-neighbor $b_1$ site.  The possible
values of $\delta$ are $-\frac{\sqrt{3}}{2}a\hat{\bx}+\frac{1}{2}a\hat{\by}$,
$\frac{\sqrt{3}}{2}a\hat{\bx}+\frac{1}{2}a\hat{\by}$, and $-a\hat{\by}$, where the
lattice constant $a\approx 1.4\text{ \AA}$.  Whenever there is a sum on $\delta$,
we sum over these three values; if $\delta$ appears without a summation over it, on
the other hand, then we choose one of these three values for it.

We may derive our low-energy effective theory for the above system by either projecting
out the high-energy modes\cite{NilssonPRB2008} or equivalently by writing the above theory as
a coherent-state path integral, integrating out the dimerized sites\cite{VafekPRB2010,ThrockmortonArXiv},
and expanding around the $\bK$ and $\bK'$ points.  The resulting theory is
\begin{eqnarray}
\mathcal{H}=\mathcal{H}_0+\mathcal{H}_{int}
\end{eqnarray}
where
\begin{eqnarray}
\mathcal{H}_0=\sum_{|\bk|<\Lambda}\sum_{\sigma=\uparrow,\downarrow}\psi^{\dagger}_{\bk\sigma}
H_{\bk}\psi_{\bk\sigma}. \label{eq:H0}
\end{eqnarray}
In the above Hamiltonian, the Fermi spinor, which describes the
modes in the vicinity of the $\pm\bK$ points and concentrated at $b$
sites in layers 1 and 2, is
\begin{eqnarray}
\psi_{\bk\sigma}= \left(\begin{array}{c} \psi^{(b_1)}_{\bK\sigma}(\bk) \\
\psi^{(b_2)}_{\bK\sigma}(\bk)\\
\psi^{(b_1)}_{-\bK\sigma}(\bk) \\
\psi^{(b_2)}_{-\bK\sigma}(\bk)
\end{array}
\right).
\end{eqnarray}
The first of the two matrices in (\ref{eq:H0}) describes the
parabolic dispersion and the second is a linear term that results in trigonal warping:
\begin{eqnarray}
H_{\bk}&=&H^{(2)}_{\bk}+H^{(tw)}_{\bk} \label {Hkin} \\
H^{(2)}_{\bk}&=&\frac{1}{2m^*}\left((k^2_x-k^2_y)\Sigma_x+2k_xk_y\Sigma_y\right)\\
H^{(tw)}_{\bk}&=&v_3\left(k_x\Lambda_x+k_y\Lambda_y\right).
\end{eqnarray}
where
\begin{eqnarray}
\Sigma_x&=&1\sigma_1,\;\;\Sigma_y=\tau_3\sigma_2\\
\Lambda_x&=&\tau_3\sigma_1,\;\;\Lambda_y=-1\sigma_2.
\end{eqnarray}

In terms of the tight-binding parameters in our lattice Hamiltonian,
the effective mass $m^*$ is
\begin{equation}
m^*=\frac{2\gamma_1}{9a^2\gamma_0^2}
\end{equation}
and the trigonal warping velocity $v_3$ is
\begin{equation}
v_3=3a\gamma_3.
\end{equation}
Experimentally\cite{Mayorov2011,VelascoNatNano2012}, $m^*\approx 0.029m_e$,
while the value that we obtain from the above formula and the
experimental values of the hopping parameters given above is $m^*\approx 0.038m_e$.
The value of the trigonal warping velocity used in fitting the experimental data\cite{Mayorov2011}
is $v_3\approx 1.41\times 10^5\text{ m/s}$, while that
obtained from the above formula is $v_3\approx 1.91\times 10^5\text{ m/s}$ (reference
\onlinecite{LemonikPRB2010} assumes a value of $v_3=10^5\text{ m/s}$).
The origin of these admittedly unimportant and small discrepancies is unclear at this time.

\subsection{Symmetry classification} The space group symmetry
operations which leave the Hamiltonian invariant at the
$\Gamma$ point form a point group $D_{3d}$. Similarly, at the $\pm
\bK$ points, the symmetry operations form a point group $D_{3}$. The
character tables\cite{TinkhamBook} of these two groups are shown below.
\begin{center}
    \begin{tabular}{ | l | l | l | l | l | l | l |}
    \hline
    ${\bf D_{3d}}$ & $E$ & 2$C_3$ & 3$C'_2$ & $i$ & 2$S_6$ & 3$\sigma_d$\\
    \hline\hline
    $A_{1_g}$ & 1 & 1 & 1 & 1 & 1 & 1 \\ \hline
    $A_{2_g}$ & 1 & 1 & -1 & 1 & 1 & -1 \\ \hline
    $E_{g}$ & 2 & -1 & 0 & 2 & -1 & 0 \\ \hline
    $A_{1_u}$ & 1 & 1 & 1 & -1 & -1 & -1 \\ \hline
    $A_{2_u}$ & 1 & 1 & -1 & -1 & -1 & 1 \\ \hline
    $E_{u}$ & 2 & -1 & 0 & -2 & 1 & 0 \\ \hline
    \end{tabular}
\end{center}
\begin{center}
    \begin{tabular}{ | l | l | l | l |}
    \hline
    ${\bf D_{3}}$ & $E$ & 2$C_3$ & 3$C'_2$\\
    \hline\hline
    $A_{1}$ & 1 & 1 & 1  \\ \hline
    $A_{2}$ & 1 & 1 & -1 \\ \hline
    $E$ & 2 & -1 & 0  \\ \hline
    \end{tabular}
\end{center}
The sixteen $4\times 4$ matrices that operate in the layer and
$\pm\bK$ valley space, can be grouped based on their transformation
properties under these group operations. We find that
\begin{eqnarray}
A_{1g}+&:&  1_4 \nonumber\\
A_{2g}-&:&\tau_3\sigma_3\nonumber\\
E_g+&:&(1\sigma_1,\tau_3\sigma_2)\nonumber\\
A_{1u}-&:&\tau_31 \nonumber\\
A_{2u}+&:& 1\sigma_3 \nonumber\\
E_u-&:& (\tau_3\sigma_1,-1\sigma_2)\nonumber\\
A_{1\bK}+&:&\tau_1\sigma_1;\tau_2\sigma_1 \nonumber\\
A_{2\bK}-&:&\tau_1\sigma_2;\tau_2\sigma_2\nonumber\\
E_{\bK}+&:&
(\tau_11,-\tau_2\sigma_3;-\tau_21,-\tau_1\sigma_3).\nonumber
\end{eqnarray}
The $\pm$ next to the name of the representation denotes whether the
particular operator is even or odd under time reversal symmetry.  An equivalent
classification can be found in Ref.\ \onlinecite{LemonikArXiv}, though
the notation is different.

\subsection{Interaction Hamiltonian}
 As shown previously, assuming the
microscopic lattice interactions are falling off faster than
$1/r^2$, there are nine marginal interaction couplings at the
Gaussian fixed point when $T=0$ and $v_3=0$. The interaction term in
the Hamiltonian is therefore
\begin{eqnarray}
\mathcal{H}_{int}&=&\frac{1}{L^2}\sum_{S}
\frac{g_S}{2}\sum_{\bk,\bk',\bq}\sum_{\sigma,\sigma'}\psi^{\dagger}_{\bk\sigma}S\psi_{\bk+\bq,\sigma}\psi^{\dagger}_{\bk'\sigma'}S\psi_{\bk'-\bq,\sigma'} \nonumber \\ \label {Hint}
\end{eqnarray}
The sum over $S$ includes the $16$ matrices belonging to the $9$
representations. Since the couplings for the squares of the
operators belonging to the same representation must be the same, we
have $9$ independent couplings. So, for example, for the $E_g$
representation, the corresponding interaction term is schematically
$\frac{1}{2}g_{E_g}\left(\left(\psi^{\dagger}_\sigma 1\sigma_1\psi_\sigma\right)^2+
\left(\psi^{\dagger}_\sigma\tau_3\sigma_2\psi_\sigma\right)^2\right)$.

We may think of $g_{E_\bK}$ as representing back scattering, $g_{A_{1g}}+g_{A_{2u}}$
as representing intralayer scattering, and $g_{A_{1g}}-g_{A_{2u}}$ as representing
interlayer scattering, as is demonstrated in our previous work\cite{ThrockmortonArXiv}.
If we introduce a density-density interaction $V(\br)$ into the microscopic tight-binding
Hamiltonian, then the forms of these couplings are given by Equations (119)-(121) in
Ref.\ \onlinecite{ThrockmortonArXiv}, where they are denoted by $g_{A_1}$, $g_{C_1}$, and
$g_\beta$, respectively.

As we will see shortly, if we start with these three couplings, then the other
six will be generated under RG.  All nine couplings may also be thought of as
interactions between local fluctuations of different order parameters.

\section{Finite-temperature Renormalization Group}
We are interested in introducing the temperature $T$ and the
trigonal warping velocity $v_3$ explicitly into our
renormalization group transformations. To proceed, we rewrite the
partition function as a coherent-state Grassman path integral:
\begin{eqnarray}
Z=\int\mathcal{D}(\psi^*,\psi)e^{-S_0-S_{int}},
\end{eqnarray}
where
\begin{eqnarray}
S_0&=&\frac{1}{\beta}\sum_{n=-\infty}^{\infty}\sum_{|\bk|<\Lambda}\sum_{\sigma=\uparrow,\downarrow}
\psi^{\dagger}_{\bk\sigma}(\omega_n)
\left(-i\omega_n+H_{\bk}\right)\psi_{\bk\sigma}(\omega_n),\nonumber
\end{eqnarray}
$n$ is an integer, and the Matsubara frequency is
$\omega_{n}=(2n+1)\pi T$.
 The interaction term is
\begin{eqnarray}
S_{int}&=&\frac{1}{2}\int_0^{\beta}d\tau\int d^2\br\sum_{S}g_S
\left(\sum_{\sigma}\psi^{\dagger}_{\sigma}(\br,\tau)
S\psi_{\sigma}(\br,\tau)\right)^2\nonumber\\
\end{eqnarray}
and
\begin{eqnarray}
\psi_{\sigma}(\br,\tau)=\frac{1}{\beta}\sum_{n=-\infty}^{\infty}\frac{1}{L}\sum_{|\bk|<\Lambda}e^{-i\omega_n\tau}e^{i\bk\cdot\br}\psi_{\bk \sigma}(\omega_n).
\end{eqnarray}
Equivalently, we may write the interaction term as
\begin{eqnarray}
S_{int}=\frac{1}{2}\int_0^{\beta}d\tau\int d^2\br\sum_{j=1}^{9}g_j\sum_{m=1}^{m_j}\left (\psi^\dag(\br,\tau)\Gamma_j^{(m)}\psi(\br,\tau)\right )^2. \nonumber \\
\end{eqnarray}
Note the absence of explicit spin subscripts on the (eight-component) $\psi=(\psi_{\uparrow},\psi_{\downarrow})^T$ fields.  The $\Gamma_j^{(m)}$ matrices are defined in Eqs. \eqref{Eq:GammaMat1}-\eqref{Eq:GammaMat9}, and $m_j$ is the multiplicity of the $j$th representation.

Our renormalization group procedure consists of splitting the $\psi$ fields
into fast and slow modes and progressively integrating out the fast modes
with momenta restricted to the small shell $\Lambda (1-\Delta \ell)<|\bk|<\Lambda$
with no restriction on the Matsubara frequencies $\omega_n$.
After each such mode
elimination, we choose to rescale the momenta in the effective
action for the slow modes such that the new cutoff is again
$\Lambda$ and that the $H^{(2)}_{\bk}$ term is left invariant. If we
also wish to keep the $i\omega_n$ term invariant, and take $\Delta\ell$
to be infinitesimal, we find that the temperature and the trigonal warping
velocity flow under RG as
\begin{eqnarray}
\frac{dT}{d\ell}&=&2T\;\;\Rightarrow\;T(\ell)=e^{2\ell}T, \\
\frac{dv_3}{d\ell}&=&v_3\;\;\Rightarrow\;v_3(\ell)=e^{\ell}v_3.
\end{eqnarray}
In general, these flow equations will be corrected once interactions
are taken into account, but for the couplings of choice here, the
corrections appear only at two-loop order.

To one-loop order, the RG flows of the coupling constants have the
form
\begin{eqnarray}\label{eq:gflow}
\frac{dg_i}{d\ell}&=&\sum_{j=1}^{9}\sum_{k=1}^{9}g_jg_k \sum_{a=1}^4
A^{(a)}_{ijk}\Phi_{a}\left(\nu_3(\ell),t(\ell)\right),
\end{eqnarray}
where $i$, like $j$ and $k$, extends over the aforementioned nine
irreducible representations of the groups of the wavevector $\Gamma$
and $\pm \bK$; $A^{(a)}_{ijk}$ are listed in
Appendix C. The dimensionless parameters that enter as the arguments
of the $\Phi$ functions are
\begin{eqnarray}
\nu_3(\ell)&=&\frac{v_3(\ell)}{\Lambda/2m^*}, \\
t(\ell)&=&\frac{T(\ell)}{\Lambda^2/2m^*}.
\end{eqnarray}
The $\Phi$ functions are determined by the integrals in Eqs. \eqref{Eq:Phi1}--\eqref{Eq:Qpm}.

As shown in Appendix \ref{App:AsympPhi}, these integrals can be evaluated
explicitly when $\nu_3=0$ in terms of elementary functions or when
$t=0$ in terms of complete elliptic integrals.

In the discussion that follows, we will make use of the asymptotic behavior in the limit
of $\ell\to\infty$:
\begin{eqnarray}
\Phi_{a}\left(\nu_3(\ell),t(\ell)\right)&=&\frac{e^{-2\ell}}{2t}+\ldots\text{ for }a=1,2, \label{Eq:PhiAsymp12} \\
\Phi_{a}\left(\nu_3(\ell),t(\ell)\right)&=&\frac{e^{-6\ell}}{12t^3}+\ldots\text{ for }a=3,4, \label{Eq:PhiAsymp34}
\end{eqnarray}
where $t=t(0)$ is the initial dimensionless temperature and ``$\ldots$'' represent terms that are smaller than the leading terms.

\subsection{General analysis of the RG flows}
In general, the flow equations (\ref{eq:gflow}) describe two
competing tendencies. The term proportional to $g_jg_k$ tends to
cause an increase of the absolute value of the coupling constants as
$\ell$ increases, while the $\Phi$ functions tend to zero as $\ell$
increases due to the increase of their arguments $\nu_3(\ell)$ and
$t(\ell)$. Numerical analysis of the flow equations reveals that,
for fixed values of the initial couplings and for a sufficiently
large value of the initial temperature $t$, there is a certain value
of $\ell$ where the flow becomes stagnant and the coupling constants
$g$ tend to {\it finite} values as $\ell\rightarrow\infty$.
Therefore, if the initial couplings are small, they remain small as
long as the initial temperature is sufficiently large even as {\it
all} the modes are integrated out. In this regime, weak-coupling RG
is entirely justified. Lowering the initial temperature, while
keeping the initial couplings fixed, causes an increase of the value
of the RG parameter $\ell$ where the coupling constants stop flowing
and an increase in the limiting value of the coupling constants. At
a critical initial temperature $t_c$, the coupling constants $g$
diverge as $\ell\rightarrow \infty$. For an initial temperature
$t<t_c$, the coupling constants diverge at finite $\ell$.

The role of trigonal warping is to cause additional suppression
of the increase of the absolute value of the coupling constants.
Thus, for fixed initial values of the coupling constants and for
sufficiently large initial $v_3$, the $g$'s do not diverge even at
$t=0$.

Therefore, as stated previously\cite{VafekYangPRB2010}, for fixed
initial $v_3$, a critical value of the initial coupling(s) must be
exceeded for a runaway flow of the coupling constant(s), which we
associate with a phase transition, to occur.

In order to understand the nature of the possible ordering tendencies,
we first analyze the asymptotic behavior of the equations
(\ref{eq:gflow}) when $t=t_c>0$ and $\ell\rightarrow \infty$.
Provided that at least one coupling $g_r$ diverges, we have managed to
enumerate {\it all} possible solutions for the stable ``rays'' along
which ratios with the other couplings $g_j/g_r$ tend to constants.
The detailed analysis of these solutions is given in Sec.\ \ref {SectionFixedRatios}.
Along such a stable ray, all nine differential equations ``collapse'' onto one, namely,
\begin{eqnarray}\label{eq:g RGasymptotics}
  \frac{\rmd g_r}{\rmd \ell}=\mathcal{A}_{(r)} g_r^2 \frac{e^{-2\ell}}{2t_c}+\ldots,\;\;\mbox{as}\;\;\ell\rightarrow\infty.
\end{eqnarray}
Here, and in the remainder of the paper, if an index is in parentheses [e.g. $(r)$], then
there is no automatic summation over $r$ unless explicitly stated.
The coefficient $\mathcal{A}_{(r)}$ depends on the stable ray along which
the couplings diverge and ``$\ldots$'' denotes terms that vanish
faster than $e^{-2\ell}$. Combining the asymptotic behavior of the $\Phi$ functions
as $\ell \to \infty$, given by Eqs. \eqref{Eq:PhiAsymp12}, \eqref{Eq:PhiAsymp34}, and \eqref {eq:gflow}, the coefficient may be expressed as
\begin {equation}
  {\mathcal A}_{(r)} = 2 \sum_{j=1}^{9}\sum_{k=1}^{9} \sum_{a=1}^{2} A_{rjk}^{(a)} \rho^{(r)}_j \rho^{(r)}_k, \label {mathcalA}
\end {equation}
where the $\rho_j^{(r)} = g_j / g_r$ is the ratio of two couplings along the stable ray.
The solution of differential equation \eqref {eq:g RGasymptotics} is
\begin{eqnarray}\label{eq:g asymptotics}
  g_r(\ell)=\frac{4t_c}{\mathcal{A}_{(r)}}e^{2\ell}+\ldots,\;\;\mbox{as}\;\;\ell\rightarrow\infty,
\end{eqnarray}
where ``$\ldots$'' denotes terms that are smaller than $e^{2\ell}$ as
$\ell\rightarrow\infty$.

\subsection{Susceptibilities and the nature of the symmetry breaking}

To find out what symmetry-breaking tendencies dominate, we start by
introducing source terms into our action:
\begin{eqnarray}
  &&\Delta S=\sum_{i=1}^{32}\Delta^{ph}_i\frac{1}{\beta}\sum_{n=-\infty}^{\infty}\sum_{\bk}
    \psi^\dag_{\bk}(\omega_n)O^{(i)}\psi_{\bk} (\omega_n) + \nonumber \\
  &&\tfrac{1}{2}\sum_{i=1}^{16}\Delta^{pp}_i\frac{1}{\beta}\sum_{n=-\infty}^{\infty}\sum_{\bk}\psi^\dag_{\bk}(\omega_n)\tilde{O}^{(i)}\psi^*_{-\bk}(-\omega_n) +\text{c.c.}\nonumber\\
\end{eqnarray}
We may think of these terms as ``forces'' that couple to
various observables, which acquire nonzero averages whenever the
system enters the appropriate phase. Note that only 18 of the 32
particle-hole source terms introduced are symmetry inequivalent.
Similarly, only nine of the 16 particle-particle source terms are
inequivalent. The transformation properties of the former under the
various symmetry group operations are summarized in Table
\ref{PH_Phases}.  Again, note the absence of explicit spin
subscripts on the (eight-component)
$\psi=(\psi_{\uparrow},\psi_{\downarrow})^T$ fields.  Terms such as
$\psi^\dag{\tilde O}\psi^*$ should be understood as matrix
multiplication, i.e., $\sum_{\alpha,\beta=1}^{8}\psi^*_\alpha{\tilde
O}_{\alpha\beta}\psi^*_\beta$.  We will see later that only two of
the particle-particle, or superconducting, orders can appear, namely
the $A_{1g}$ and $A_{2u}$ orders.  These correspond to $s_{++}$- and
$s_{+-}$-wave superconducting orders, respectively.  Both are
$s$-wave, but the $s_{++}$ order parameter has the same sign on both
layers, while the $s_{+-}$ order has opposite signs on each layer.
\begin{table*}
\centering
\begin{tabular}{ | c | c | c | c | c | c | c | c |}
\hline
Group rep. & Matrices & Trans. & TRS & Inv. & Mirror refl. ($\sigma_d$) &  \\
\hline
\hline $A_{1g}$ charge & $1_4\otimes 1$ & $+$ & $+$ & e & e & Charge instability \\
\hline $A_{2g}$ charge & $\tau_3\sigma_3\otimes 1$ & $+$ & $-$ & e & o & Anomalous quantum Hall\cite{HaldanePRL1988,NandkishorePRB2010} \\
\hline $E_g$ charge & $(1\sigma_1, \tau_3\sigma_2)\otimes 1$& $+$ & $+$ & e & e/o & Nematic\cite{VafekYangPRB2010,LemonikPRB2010} \\
\hline $A_{1u}$ charge & $\tau_31\otimes 1$ & $+$ & $-$ & o & o & Loop current\cite{ZhuArXiv} \\
\hline $A_{2u}$ charge & $1\sigma_3\otimes 1$ & $+$ & $+$ & o & e & Layer-polarized\cite{MinPRB2008,NandkishorePRL2010} \\
\hline $E_u$ charge & $(\tau_3\sigma_1,-1\sigma_2)\otimes 1$ & $+$ & $-$ & o & o/e & Loop current II (ME2) \\
\hline $A_{1\bK}$ ($A_{1g}$/$A_{1u}$) charge & $\tau_1\sigma_1\otimes 1; \tau_2\sigma_1\otimes 1$ & $-$ & $+$ & e/o & e/o & Kekul\'e\cite{HouPRL2007} \\
\hline $A_{2\bK}$ ($A_{2u}$/$A_{2g}$) charge & $\tau_1\sigma_2\otimes 1;\tau_2\sigma_2\otimes 1$ & $-$ & $-$ & o/e & e/o & Kekul\'e current \\
\hline $E_\bK$ ($E_g$/$E_u$) charge & $(\tau_1 1, -\tau_2\sigma_3)\otimes 1; (-\tau_2 1, -\tau_1\sigma_3)\otimes 1$ & $-$ & $+$ & e/o & (e/o)/(o/e) & Charge density wave \\
\hline
\hline $A_{1g}$ spin & $1_4\otimes\vec{\sigma}$ & $+$ & $-$ & e & e & Ferromagnetic \\
\hline $A_{2g}$ spin & $\tau_3\sigma_3\otimes\vec{\sigma}$ & $+$ & $+$ & e & o & Quantum spin Hall\cite{ThrockmortonArXiv,SchererArXiv,LemonikArXiv} \\
\hline $E_g$ spin & $(1\sigma_1,\tau_3\sigma_2)\otimes\vec{\sigma}$ & $+$ & $-$ & e & e/o & Spin nematic \\
\hline $A_{1u}$ spin & $\tau_3 1\otimes\vec{\sigma}$ & $+$ & $+$ & o & o & Staggered spin current \\
\hline $A_{2u}$ spin & $1\sigma_3\otimes\vec{\sigma}$ & $+$ & $-$ & o & e & Layer AF\cite{CastroNetoRMP2009,VafekPRB2010,KharitonovArXiv} \\
\hline $E_u$ spin & $(\tau_3\sigma_1, -1\sigma_2)\otimes\vec{\sigma}$ & $+$ & $+$ & o & o/e & Loop spin current II \\
\hline $A_{1\bK}$ ($A_{1g}$/$A_{1u}$) spin & $\tau_1\sigma_1\otimes\vec{\sigma};\tau_2\sigma_1\otimes\vec{\sigma}$ & $-$ & $-$ & e/o & e/o & Spin Kekul\'e \\
\hline $A_{2\bK}$ ($A_{2u}$/$A_{2g}$) spin & $\tau_1\sigma_2\otimes\vec{\sigma}; \tau_2\sigma_2\otimes\vec{\sigma}$ & $-$ & $+$ & o/e & e/o & Spin Kekul\'e current \\
\hline $E_\bK$ ($E_g$/$E_u$) spin & $(\tau_1 1, -\tau_2\sigma_3)\otimes\vec{\sigma}; (-\tau_2 1, -\tau_1\sigma_3)\otimes\vec{\sigma}$ & $-$ & $-$ & e/o & (e/o)/(o/e) & Spin density wave \\
\hline
\end{tabular}
\caption{\label{PH_Phases}Table of all particle-hole phases considered, listed according to what representation of the $D_{3d}$ point group they transform.  The Kekul\'e and density waves have a wave vector of $\bK$.}
\end{table*}
To one-loop order, we find
\begin{eqnarray}\label{eq:Delta RG eqs}
  \frac{\rmd \ln\Delta^{ph}_i}{\rmd \ell}&=&2+\sum_{j=1}^{9}\sum_{a=1}^4B^{(a)}_{ij}g_j(\ell)\Phi_a\left(\nu_3(\ell),t(\ell)\right), \label {dlogDeltaph}\\
  \frac{\rmd \ln\Delta^{pp}_i}{\rmd \ell}&=&2+\sum_{j=1}^{9}\sum_{a=1}^4\tilde{B}^{(a)}_{ij}g_j(\ell)\Phi_a\left(\nu_3(\ell),t(\ell)\right), \label {dlogDeltapp}
\end{eqnarray}
where the ($32\times 9$) matrix $B^{(a)}_{ij}$ and the ($16\times 9$) matrix $\tilde B^{(a)}_{ij}$ are defined by Eqs. \eqref{Eq:BDef}--\eqref{Eq:B342} and
\eqref{Eq:Bt12}--\eqref{Eq:Bt34}.  Note that Eqs. \eqref{dlogDeltaph} and \eqref{dlogDeltapp} can be readily integrated, and we find that
\begin{equation}\label{eq:Delta_integrated}
\Delta_i^{ph/pp}(\ell)=\Delta_i^{ph/pp}(0)e^{2\ell}\exp[\Omega_i^{ph/pp}(\ell)],
\end{equation}
where
\begin{eqnarray}
\Omega_i^{ph}(\ell)&=&\sum_{j=1}^{9}\sum_{a=1}^4B^{(a)}_{ij}\int_{0}^{\ell}d\ell'\,g_j(\ell')\Phi_a\left(\nu_3(\ell'),t(\ell')\right), \nonumber \\ \\
\Omega_i^{pp}(\ell)&=&\sum_{j=1}^{9}\sum_{a=1}^4{\tilde B}^{(a)}_{ij}\int_{0}^{\ell}d\ell'\,g_j(\ell')\Phi_a\left(\nu_3(\ell'),t(\ell')\right). \nonumber \\
\end{eqnarray}

At $t=t_c>0$, as $\ell\rightarrow\infty$ the $e^{2\ell}$ increase of a divergent coupling
$g_r$ exactly balances the $e^{-2\ell}$ decrease of the $\Phi$ functions and the
right hand sides of the above equations tend to constants,
\begin{eqnarray}
\frac{d\ln\Delta^{ph}_i}{d\ell}&=&2+\frac{2\mathcal{B}^{ph}_{i(r)}}{\mathcal{A}_{(r)}}\;\;\mbox{as}\;\ell\rightarrow\infty,\nonumber\\
\frac{d\ln\Delta^{pp}_i}{d\ell}&=&2+\frac{2\mathcal{B}^{pp}_{i(r)}}{\mathcal{A}_{(r)}}\;\;\mbox{as}\;\ell\rightarrow\infty.
\label{eq:B ph and pp (etas)}
\end{eqnarray}
In other words, the engineering dimensions of the source terms, which
are equal to $2$, are corrected by the anomalous dimensions
\begin {equation}
  \eta^{ph/pp}_i = \frac { 2 {\mathcal B}^{ph/pp}_{i(r)} }{ {\mathcal A}_{(r)}} \label{eq:AnomDim}
\end {equation}
due to the electron-electron interactions.
Again, in the above equation, there is no summation over $r$, which corresponds to the divergent coupling $g_r$ that we divided by.
The values of the ${\mathcal B}$'s are
\begin {eqnarray}
  {\mathcal B}^{ph}_{i(r)} &=& 2 \sum_{k=1}^{9} (B^{(1)}_{ik} + B^{(2)}_{ik}) \rho^{(r)}_k, \label{eq:BCoeffPH} \\
  {\mathcal B}^{pp}_{i(r)} &=& 2 \sum_{k=1}^{9} (\tilde B^{(1)}_{ik} + \tilde B^{(2)}_{ik}) \rho^{(r)}_k, \label{eq:BCoeffPP}
\end {eqnarray}
where $B^{(1/2)}$ is given by the sum of Eqs. \eqref{Eq:B121} and \eqref{Eq:B122} and ${\tilde B}^{(1/2)}$ is given by \eqref{Eq:Bt12}.  Note that the expressions for ${\mathcal A}_{(r)}$ and $\mathcal B^{ph/pp}_{(r)}$ depend on the choice of $g_r$,
but the $\eta^{ph/pp}_i$'s do not.

In order to calculate the physical susceptibility toward various
ordering tendencies, we calculate the correction to the free energy
due to the presence of the symmetry breaking source terms\cite{Nelson1975}. We find
that
\begin{eqnarray}
&&\delta f(\Delta)=  \label{eq:free energy} \\
&&-\frac{\ms}{16\pi}\sum_{i=1}^{32}\int_{0}^{\infty}\rmd\ell\,e^{-4\ell}[\Delta_{i}^{ph}(\ell)]^2\sum_{a=1}^{4}\alpha_{a,i}^{ph}\Phi_{a}(\nu_3(\ell),t(\ell)) \nonumber \\
&&-\frac{\ms}{16\pi}\sum_{i=1}^{16}\int_{0}^{\infty}\rmd\ell\,e^{-4\ell}|\Delta_{i}^{pp}(\ell)|^2\sum_{a=1}^{4}\alpha_{a,i}^{pp}\Phi_{a}(\nu_3(\ell),t(\ell)), \nonumber
\end{eqnarray}
The $\alpha$ coefficients are given in Appendix \ref{App:Coeff_FE} by Eqs. \eqref{Eq:alpha12ph}-\eqref{Eq:alpha34pp}.

The susceptibilities are then simply given by second derivatives of the free energy with
respect to the bare values of the appropriate source terms,
\begin{eqnarray}
\chi^{ph}_i&=&-\frac{\partial^2 f}{\partial[\Delta_i^{ph}(\ell=0)]^2}, \\
\chi^{pp}_i&=&-\frac{\partial^2 f}{\partial[\mbox{Re}\,\Delta_i^{pp}(\ell=0)]^2}= -\frac{\partial^2 f}{\partial[\mbox{Im}\,\Delta_i^{pp}(\ell=0)]^2}. \nonumber \\
\end{eqnarray}
Using Eqs.\ \eqref{eq:Delta_integrated} and \eqref{eq:free energy}, we find that the susceptibilities given above may be written as
\begin{eqnarray}
\chi^{ph/pp}_i = \frac{\ms}{8\pi} \sum_{a=1}^{4}\alpha_{a,i}^{ph/pp}
  \int_{0}^{\infty}{\rmd}\ell ~ e^{2 \Omega_i^{ph/pp}(\ell)} \Phi_{a}(\nu_3(\ell),t(\ell)). \nonumber \\ \label {chifinal}
\end{eqnarray}
Note that the source terms, $\Delta_i^{ph/pp} (\ell = 0)$, being auxiliary fields, do not appear.

Any divergence in the susceptibilities has to come from the regions of
large $\ell$ in Eq.\ \eqref {eq:free energy} where the asymptotic expressions derived earlier hold.
Therefore, since, for $t=t_c>0$, the asymptotic behavior of the $\Phi$ functions
is $e^{-2\ell}$, the condition for the divergence of a susceptibility
in a particle-hole or particle-particle channel $i$ is
\begin{eqnarray}\label{eq: div susc}
  \eta_i^{ph/pp} >1.
\end{eqnarray}

Next, we will relate the anomalous dimensions of the source terms $\eta_i^{ph/pp}$
to the susceptibility exponents $\gamma_i^{ph/pp}$.

\subsection{Susceptibility exponents}

For $t>t_c$, sufficiently close to $t_c$ the asymptotic behavior of
the coupling constants is still approximately described by Eq.\
\eqref{eq:g RGasymptotics}. If we integrate it from $\ell_0$ to
$\ell$, both of which are asymptotically large (and temperature
independent), but not infinite, then we find
\begin{eqnarray}
\frac{1}{g_r(\ell,t)}&=&\frac{1}{g_r(\ell_0,t)}-\frac{\mathcal{A}_{(r)}}{4t}\left(e^{-2\ell_0}-e^{-2\ell}\right).
\end{eqnarray}
At $t_c$ we have $1/g_r(\ell_0,t_c)= {\mathcal A}_{(r)} e^{-2\ell_0}/4t_c$ and
we can write the above equation as
\begin{eqnarray}
\frac{1}{g_r(\ell,t)}&=&\left(\frac{1}{g_r(\ell_0,t)}-\frac{1}{g_r(\ell_0,t_c)}\right) \\
&-&\left(\frac{1}{t}-\frac{1}{t_c}\right)\left(\frac{\mathcal{A}_{(r)}}{4}e^{-2\ell_0}\right)
+\frac{\mathcal{A}_{(r)}}{4t}e^{-2\ell}. \nonumber
\end{eqnarray}
Since $\ell_0$ is finite, $g_r(\ell_0,t)$ is analytic in $t$ at $t_c$
and can be expanded as
\begin{eqnarray}
g_r(\ell_0,t)\approx g_r(\ell_0,t_c)+(t-t_c)\frac{\partial}{\partial
t}g_r(\ell_0,t)\bigg|_{t_c}+\ldots,
\end{eqnarray}
where ``$\ldots$'' represents terms of order $(t-t_c)^2$ and higher.
Therefore
\begin{eqnarray}
g_r(\ell,t)\approx\frac{1}{c_r (t-t_c)+\frac{\mathcal{A}_{(r)}}{4t}e^{-2\ell}},\;\;\mbox{as}\;\ell\rightarrow\infty,\;t\rightarrow
t^+_c\end{eqnarray} where
\begin{eqnarray}
c_r&=&\frac{\partial}{\partial t}\frac {1}
{g_r(\ell_0,t)} \bigg|_{t_c}+\frac{\mathcal{A}_{(r)}}{4t^2_c}e^{-2\ell_0}.
\end{eqnarray}
Note that $c_r \mathcal{A}_{(r)}>0$ since $g_r(\ell_0,t)$ increases in
magnitude as $t\rightarrow t^+_c$.

The flow of the source terms at large $\ell$ at $t>t_c$ is
determined by substituting the above result into the Eqs.\
\eqref{dlogDeltaph} - \eqref{dlogDeltapp} and taking the asymptotic
limit of the $\Phi$ functions at large $\ell$:
\begin{eqnarray}
\frac{d\ln\Delta^{ph}_i}{d\ell}&=&2+\frac{\mathcal{B}^{ph}_{i(r)}}{2t}\frac{e^{-2\ell}}{c_r (t-t_c)+\frac{\mathcal{A}_{(r)}}{4t}e^{-2\ell}}\\
\frac{d\ln\Delta^{pp}_i}{d\ell}&=&2+\frac{\mathcal{B}^{pp}_{i(r)}}{2t}\frac{e^{-2\ell}}{c_r (t-t_c)+\frac{\mathcal{A}_{(r)}}{4t}e^{-2\ell}}.
\end{eqnarray}
Integrating from $\ell_0$ to $\ell$ and substituting to
Eq.(\ref{eq:free energy}), we find that the singular contribution to
the susceptibility for the symmetry breaking source term $\Delta_i$
is
\begin{eqnarray}
\chi_i^{ph/pp}&\approx& \left(t-t_c\right)^{-\gamma^{ph/pp}_i}
\end{eqnarray}
where
\begin{eqnarray}
\gamma_{i}^{ph/pp}&=&\eta^{ph/pp}_i-1. \label{SuscExp_Formula}
\end{eqnarray}
Clearly, the susceptibility for a particular order diverges if the
condition (\ref{eq: div susc}) is satisfied. Note that only if
$\eta_i^{ph/pp}=2$ do the susceptibility exponents
acquire their mean-field values. This is in general not the case
here, as will be elaborated on in the next section.

It is also important to stress that these exponents are obtained
within the one-loop approximation {\it of the fermionic theory} and
are therefore not expected to be accurate. They are also not
expected to be equal to the one-loop exponents obtained within an
$\eps$-expansion of the corresponding bosonic theory, with the Landau
functional for the ordering field. The ultimate critical behavior is
determined by the universality class of such a bosonic theory. As an
example, the finite temperature phase transition into the nematic
state belongs to the two dimensional three-state Potts model \cite {VafekYangPRB2010}
universality class for which the exponent $\gamma=13/9$ (see Ref.\ \onlinecite{WuRMP1982}). However,
within our one-loop fermionic RG treatment, $\gamma$ does not exceed $2/3$.
Nevertheless, the exponents calculated within the present
approximation give us important information about the physical
character of the dominant ordering tendency, without any {\it a priori}
bias toward any given order.

Next, we will explicitly calculate the RG flows using numerical
integration of the RG equations for the couplings, the symmetry-breaking
vertex terms and the physical susceptibilities. We do so
for any interaction that can be written as
$\sum_{\bq} \sum_{i,j}V_{ij}(\bq) n_i (\bq) n_j (-\bq)$,
where $V_{ij} (\bq)$ is finite for any $\bq$; $i$ and $j$ run over sublattice and layer indices.
To leading order in small $V$,
$\bq$ is either near $0$ or near $\pm2\bK$. Such a microscopic lattice interaction will
initially lead to only three of the nine four-fermion coupling constants in
the low-energy effective field theory being finite \cite {ThrockmortonArXiv}, i.e., $g_{A_{1g}}\big|_{\ell=0}\neq0$,
$g_{A_{2u}}\big|_{\ell=0}\neq0$, $g_{E_{K}}\big|_{\ell=0}\neq0$.

\begin{figure}[h]
\begin{center}
\includegraphics[width=0.49\textwidth]{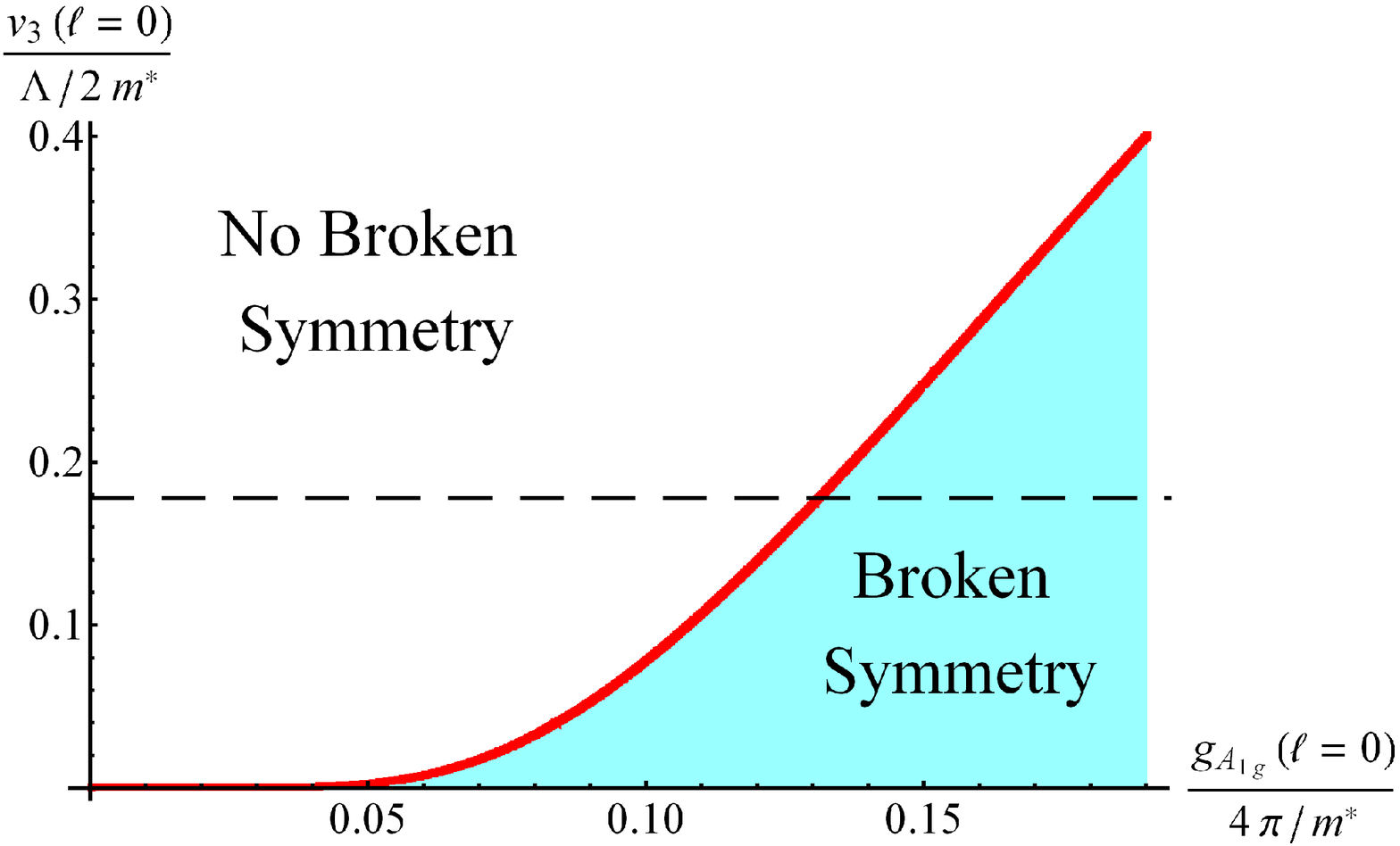}
a)
\includegraphics[width=0.49\textwidth]{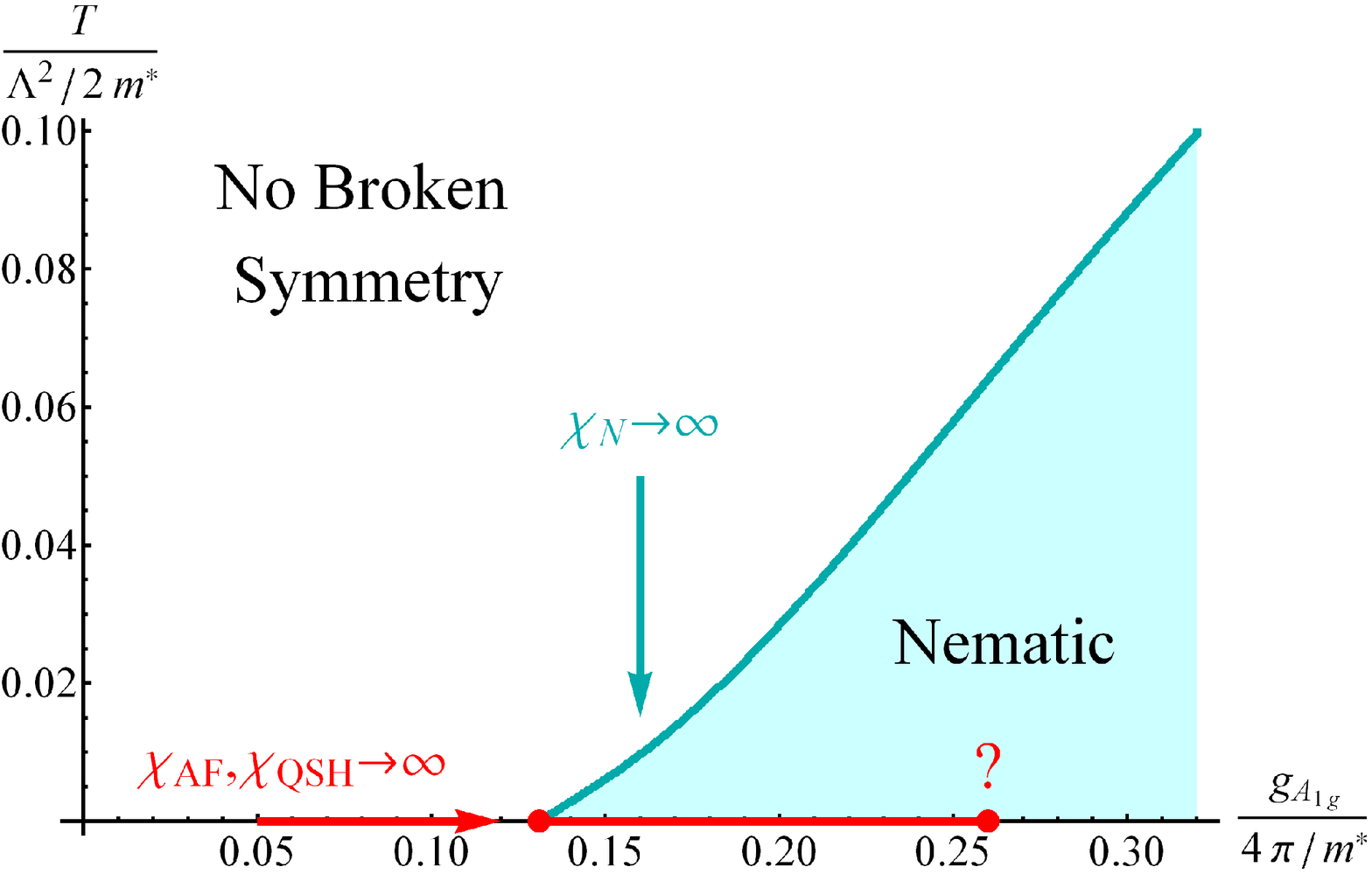}
b)
\end{center}
  \caption{(Color online) Phase boundaries for bilayer graphene with forward scattering only. (a) At finite trigonal warping,
    $\nu_3 = 2 m^* v_3 /\Lambda$ and $T=0$, the
    bare $g_{A_{1g}}$ must exceed a critical value, given by the red line, in order for the
    system to enter a broken symmetry phase. Along the dashed line, $\nu_3=0.178$,
    which is the value used to fit the experimental data in Ref.\ \onlinecite {Mayorov2011}. (b) The transition
    temperature as a function of the initial value of $g_{A_{1g}}$ at $\nu_3=0.178$. At
    any finite $t_c$, the only susceptibility that diverges corresponds to the nematic order parameter
   ($E_g$ charge), as shown in Fig.\ \ref {FigSusceptibilityNem}. }\label{fig:Crit v3 for gA1g}
\end{figure}

\subsection{Forward scattering limit: nematic} \label {SecFwdNematic}

\begin{figure}[h]
\begin{center}
\includegraphics[width=0.49\textwidth]{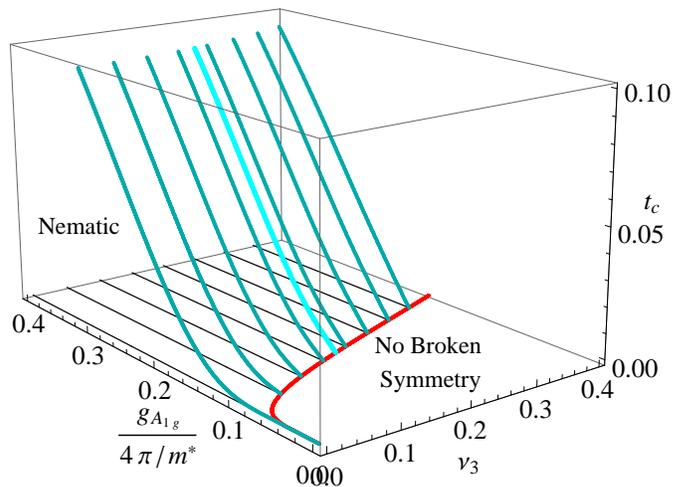}
\end{center}
  \caption{(Color online) The phase diagram for different values of the forward scattering coupling, $g_{A_{1g}}$,
    trigonal warping velocity, $\nu_3 = 2 m^* v_3 / \Lambda$, and temperature $t= 2m^* T / \Lambda^2$.
    At any finite $t_c$ and at any value of $\nu_3$ the nematic susceptibility
    is the only one that diverges. The lighter cyan line, also shown in Fig.\ \ref {fig:Crit v3 for gA1g}(b),
    corresponds to $\nu_3=0.178$, which is the value used throughout
    the paper. }\label{Figgv3Tin3D}
\end{figure}

For $V_{ij} (\pm 2\bK)=0$ and equal inter- and
intralayer interactions, the only  non-zero bare interaction  is
$g_{A_{1g}}$. Without scattering between the $\bK$
and $\bK'$ valleys at the microscopic level, no new scattering between them
can be generated in the RG flow, Eq.\ \eqref {eq:gflow}. In other words,
$g_{A_{1\bK}}$, $g_{A_{2 \bK}}$, and $g_{E_\bK}$ remain zero.
The only couplings that appear in this case are those in the $g$ and $u$ representations.
We studied the problem numerically and present the main results in Figs.\ \ref {fig:Crit v3 for gA1g}
and \ref {Figgv3Tin3D}.

For any fixed initial $v_3$, we find that there is a critical value
of $g_{A_{1g}}$ below which the weak-coupling RG converges and no
phase transition occurs, even at $T=0$, as shown in Fig.\ \ref
{fig:Crit v3 for gA1g}a. In this phase, where no symmetry is broken,
there are four Dirac points in the vicinity of each $\bK$ point,
three of which are anisotropic and one, centered at $\bK$ or $\bK'$, which is
isotropic.

\begin{figure}[h]
\begin{center}
\includegraphics[width=0.49\textwidth]{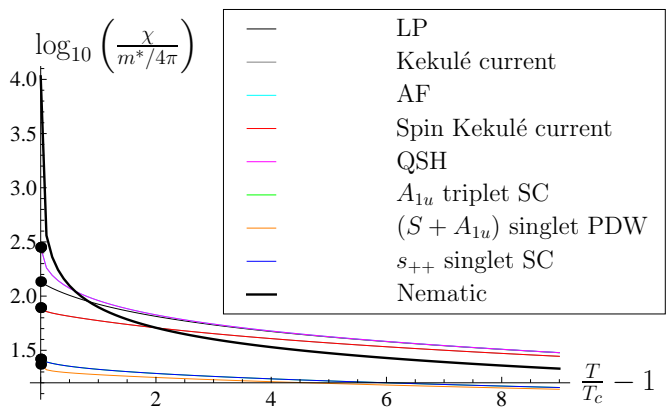}
\end{center}
  \caption{(Color online) Various susceptibilities calculated from the free energy given
    by Eq.\ \eqref {eq:free energy}  with forward scattering only. Although the nematic
    susceptibility is lower than the others at higher temperatures, it is the only susceptibility that diverges as
    the temperature is lowered towards $t_c>0$.  Here, $\nu_3=0.178$, the bare couplings are
    $g_{A_{1g}}=0.161\times 4\pi/m^*$, and all others zero.  In this case,
$t_c=0.01$.}\label{FigSusceptibilityNem}
\end{figure}

Above the critical value of $g_{A_{1g}}$,  the {\it only} susceptibility that
diverges as $t \to t_c^+ >0$ is toward an electronic nematic, i.e.,
toward a spin-singlet order parameter that transforms according to the
$E_g$ representation (see Table \ref {PH_Phases}). We therefore conclude that, immediately below this temperature,
the system enters this symmetry-breaking phase. In Ref.\
\onlinecite{Mayorov2011} the experimental data is fitted using a value of the trigonal
warping velocity corresponding to our dimensionless parameter
$\nu_3=0.178$. Here, and in the remainder of
the paper, we use this value.
The phase boundary for that particular choice of $\nu_3$ is shown in
Fig.\ \ref {fig:Crit v3 for gA1g}(b). In Fig.\ \ref
{FigSusceptibilityNem}, we show the susceptibilities as a function of temperature
for various order parameters. This plot corresponds to the
vertical arrow in Fig.\ \ref {fig:Crit v3 for gA1g}(b) where only the
nematic susceptibility diverges. We find that the susceptibility
toward the nematic order parameter, despite being smaller than the
others at large $T$, outgrows all competing susceptibilities as the
temperature is lowered toward $T_c$, and is the only susceptibility
to diverge at $T_c$.

While our analysis of the susceptibilities reveals that at $t_c>0$ only the
nematic susceptibility diverges for any fixed $\nu_3$, this susceptibility does
not diverge when approaching the
critical $g_{A_{1g}}$ from below exactly at $t=0$. Instead, we
find that the susceptibilities for the layer antiferromagnet (AF) and  quantum spin Hall
(QSH) order parameters diverge with equal exponents.
This suggests that, at $0<t<t_c$, the system orders into a nematic
state, while at $t=0$ there may be a coexistence of this state with AF and/or QSH \cite{MWargument}.

The complete phase diagram for different values of $g_{A_{1g}}$, $\nu_3$,
and $t_c$ is shown in Fig.\ \ref {Figgv3Tin3D}. For the entire range of $\nu_3$'s
shown, we always find the nematic as the leading instability at finite temperature.

In order to facilitate the comparison with the previous work, which
also deals with the forward scattering limit at $T=v_3=0$, we first
note that the three couplings $g_1$, $g_2$, and $g_3$ in
Ref.\ \onlinecite{VafekYangPRB2010} correspond to $g_{A_{1g}}$,
$g_{A_{2g}}$, and $g_{E_g}$, respectively. Because $v_3=0$, none of
the other nine four fermion couplings are generated under RG. The
``susceptibilities'' calculated therein correspond to the logarithmic
prefactors on the right-hand sides of Eqs.\ (15) and (16) in
Ref.\ \onlinecite{VafekYangPRB2010}, and are analogous to the more
general expressions in Eqs.\ \eqref{dlogDeltaph}--\eqref{dlogDeltapp}
of this publication. The physical susceptibilites, considered in
this paper, can be straightforwardly obtained from such expressions
by substituting the flow of the source terms into the formula
(\ref{eq:free energy}). At
$T=0$, the divergences appear at finite $\ell=\ell^*$, which can be set
as the upper limit on the integrals in Eq.\ \eqref{eq:free energy}. The coupling
constant ratio\cite{VafekYangPRB2010} $g_{A_{1g}}/g_{E_g}\rightarrow
0$ as $\ell\rightarrow \ell^*$. The ratio $g_{A_{2g}}/g_{E_g}$ can
approach either $m_1\approx -0.525$ or $m_3\approx 13.98$, i.e., the
minimal or the maximal root of the cubic equation $(x-14) x^2+4=0$.

The analysis of the physical susceptibilities for the conditions
stated in Ref.\ \onlinecite{VafekYangPRB2010} reveals that, as
$\ell\rightarrow \ell^*$, the {\em only} physical susceptibility
that diverges when $g_{A_{2g}}/g_{E_g}\rightarrow m_1$ is towards
the nematic; the others remain finite at $\ell^*$. Similarly, the
{\em only} susceptibility that diverges when
$g_{A_{2g}}/g_{E_g}\rightarrow m_3$ is toward the quantum anomalous
Hall state (QAH). Very recently, Fan Zhang {\it et al.} posted a
preprint \cite{FanZhangArXiv2012} in which they recovered the
$T=v_3=0$ flow equations for the three couplings in Ref.\
\onlinecite{VafekYangPRB2010}. Under equivalent conditions to those
in Ref.\ \onlinecite{VafekYangPRB2010}, they claim to have
calculated susceptibilities and that the ``strongest divergence
occurs for the flavor spin channel broken inversion symmetry.''
These results are at odds with our findings.  We believe the
discrepancy to originate from their Eq. (24), which does not
properly account for renormalization of composite operators (see
Refs. \onlinecite{PeskinSchroederBook,CardyBook}).

\subsection{General density-density interaction} \label{Sec:DDInt}

In the previous section, we have shown how a system with forward scattering only
at the bare level orders into the nematic state at any finite temperature. In general, however, other
four-fermion coupling constants may be non-zero as well. In a previous work on this
subject \cite {ThrockmortonArXiv}, two of us showed how to find the bare interaction strengths
corresponding to a screened interaction in the weak coupling limit. In addition to $g_{A_{1g}}$, two other couplings,
$g_{A_{2u}}$ and $g_{E_\bK}$, appear at $\ell=0$. Due to the presence of these couplings,
all $\beta$ functions are non-zero and all nine couplings allowed by symmetry are generated in the RG flow.

Instead of seeking a critical temperature for a given set of initial
couplings, we invert the procedure by fixing the transition
temperature to $t_c = 0.01$. This value is in accordance with the
experimentally observed symmetry-breaking energy scale of $\sim 2$
meV. We then determine what values of the initial couplings would
make the RG flow divergent at this temperature. This set of points
defines a two-dimensional surface in the three-dimensional space of
initial $(g_{A_{1g}}, g_{A_{2u}}, g_{E_\bK})$. For each point on
this surface, we find the list of phases for which the
susceptibility divergence criterion, Eq.\ \eqref {eq: div susc}, is
satisfied. For certain initial conditions it happens that two or
more susceptibilities diverge at $t_c$. In such situations we list
all the phases with diverging susceptibilities (e.g., ``N+AF''
represents the region of initial couplings where both $\chi_N$ and
$\chi_{AF}$ diverge, although not necessarily with the same
exponent). Because our formalism is valid only for $t \ge t_c$, the
resulting state may be either one of the listed phases or a
coexistence of several of these phases. In order to decide which
phase(s) is present, one needs to construct a theory valid below
$t_c$, such as a Landau theory with multiple order parameters. This
is beyond the scope of the present paper.

\begin{figure}
  \includegraphics[width=0.49\textwidth]{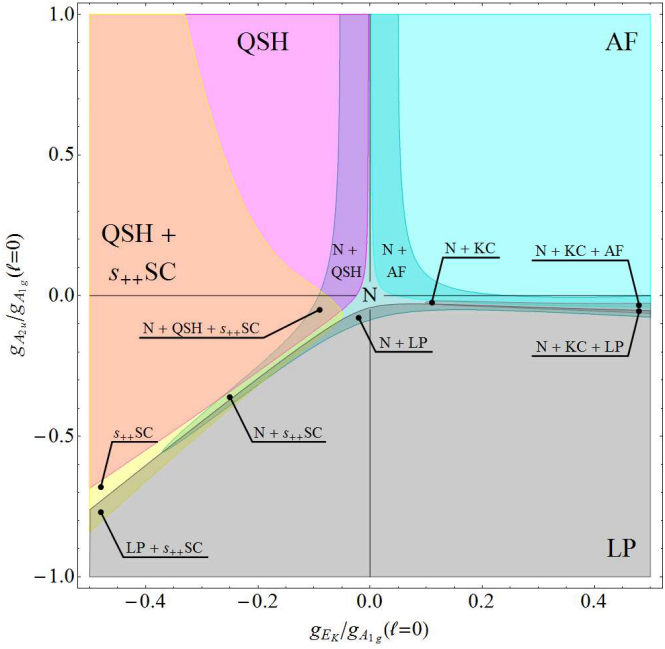}
  \caption {(Color online) The phase diagram of bilayer graphene with trigonal warping. The transition temperature is fixed to $t_c=0.01$ and
    $\nu_3 = 0.178$. Predominantly forward scattering favors the nematic (N) phase. When backscattering
    and/or the difference between inter- and intralayer scattering is considerable at the
    bare level we find other phases: the layer antiferromagnet (AF),
    the layer-polarized state (LP), the quantum spin Hall state
    (QSH),  the $s_{++}$ superconducting state ($s_{++}$ SC), and the  Kekul\'e current
    state (KC). In regions where two or more susceptibilities diverge
    at the same $t_c$ we use ``+'' to denote a ``coexistence'' of multiple
    possible phases. Whether the listed phases truly coexist or one of them is preferred must
    be determined from the full Landau function. Such an analysis is beyond the scope of this paper. } \label {FigPhaseDiagramTv3}
\end{figure}

The phase diagram we find is presented in Fig.\ \ref {FigPhaseDiagramTv3}. One should understand the axes on
this plot as follows. When the microscopic interaction has a long range, the bare values of
$g_{A_{2u}}$ and $g_{E_\bK}$ are negligible relative to $g_{A_{1g}}$. They become larger
for interactions with shorter range \cite {ThrockmortonArXiv}. For monotonically-decreasing repulsive
interaction potentials, these two couplings do not exceed $g_{A_{1g}}$ and $g_{A_{1g}}/2$,
respectively. $g_{A_{2u}}$ and $g_{E_\bK}$
reach these values in the Hubbard limit, where the only microscopic interaction term is on-site.
We therefore restrict our phase diagram to positive initial values
of $g_{A_{1g}}$, to $|g_{E_\bK}|/g_{A_{1g}}\le 1/2$, and to $|g_{A_{2u}}|/g_{A_{1g}}\le 1$.

In the given range of initial couplings, we find a rich phase diagram with the following phases:

  {\it (a) Nematic (N)}: This phase is stable for predominantly forward scattering, i.e., when both $g_{A_{2u}}$ and $g_{E_\bK}$ are
    small at the bare level. If one of these couplings remains small and the other becomes comparable to $g_{A_{1g}}$ the nematic
    susceptibility is still divergent, although other susceptibilities will diverge at these initial values
    as well. The nematic state is gapless, but it reconstructs the low-energy spectrum such that
    two out of four Dirac cones in each valley become gapped.
	
  {\it (b) Layer antiferromagnet (AF)}: This phase occurs
    when all three bare couplings are repulsive and comparable, corresponding to a short-range
    repulsive interaction. In this state the magnetization on each undimerized site is finite, with the magnetization
    within one layer pointing in one direction, and that in the other layer in the opposite direction.
	
  {\it (c) Layer-polarized state (LP)}: This phase is preferred when the
    interlayer repulsion is stronger than the intralayer repulsion ($g_{A_{2u}} (\ell= 0)
    \lesssim 0$), and the backscattering is either repulsive or weakly
    attractive ($g_{E_\bK}  (\ell= 0) \gtrsim g_{A_{2u}}  (\ell= 0)$). In this
    phase, which is gapped, there is an imbalance of the electron occupation number between the two layers. One layer is
    more occupied and the opposite layer is equally less occupied with respect to the
    symmetric, high-temperature, state.
	
  {\it (d) Quantum spin Hall state (QSH)}: This state is preferred when the backscattering is attractive  ($g_{E_\bK} (\ell= 0)
    \lesssim 0$), but if $g_{A_{2u}}$ is attractive as well, it must be weaker ($g_{E_\bK}  (\ell= 0) \lesssim g_{A_{2u}}  (\ell= 0)$).
    In this state, which is gapped, there is a spin current around each plaquette circulating in the same direction
    in both layers.
	
  {\it (e) $s_{++}$ superconductor ($s_{++}$ SC)}: The conditions for this phase are similar to the previous one
    in that the backscattering must be attractive, but it must also be roughly
    stronger than (attractive or repulsive) $|g_{A_{2u}}|$ at $\ell = 0$. This state opens a superconducting gap in both
    layers with the same sign on each layer.
	
  {\it (f) Kekul\'e current phase (KC)}: This phase appears in a thin sliver of initial couplings for
    which backscattering is repulsive and comparable
    to $g_{A_{1g}}$, while the inter- and intralayer couplings are roughly
    the same ($g_{A_{2u}} (\ell = 0) \approx 0$). It breaks lattice translational symmetry and time-reversal
    symmetry. In this phase a supercell, consisting of three regular unit cells, is formed. Within the supercell,
    two plaquettes carry  a circulating current, both in the same direction. This phase is gapped.

For a graphical illustration of some of these phases, see Fig.\ 2 in Ref.\ \onlinecite {LemonikArXiv}.

In the entire plot the values of the bare dimensionless couplings $m^* g_i / 4 \pi$
for which the system orders at the preset $t_c=0.01$ and
$\nu_3=0.178$ are always smaller than 0.15, which justifies our weak-coupling approach.

The situation does not change qualitatively with variations in temperature or in the
absence of trigonal warping --- we explored a range of temperatures $0.005 \le t_c \le 0.02$
with and without trigonal warping and found that the general structure of the phase diagram in Fig.\ \ref {FigPhaseDiagramTv3}
does not change appreciably. This is not a coincidence and, later in the paper,
we will map the phases that we may obtain
by analyzing the behavior of the flow equations in the large $\ell$ limit, where trigonal warping is irrelevant.

\subsection {The Hubbard model --- ``hidden'' symmetry} \label{Sec:Hubbard}

As an important check, we apply our RG procedure to a special case, namely the Hubbard
model, about which we already know certain exact properties. At
half filling, this model has a dynamical SO(4) symmetry \cite {CNYangSCZhangMPLB1990}
on a bipartite lattice.
This symmetry is present regardless of the sign of $U$ and the dimensionality. When $U$ is negative,
this symmetry is particularly useful because the electrons have a tendency toward pairing.  One good variational
ground state for the negative $U$ Hubbard model on a square lattice is a charge density wave, where one sublattice has a
higher occupation number than the other.  Another ground state that exhibits pairing is the $s$-wave superconductor.
The pseudospin symmetry rotates between these states. At half filling the tendency towards the charge density wave
order must therefore be the same as the tendency towards the $s$-wave superconducting order.

In the case of bilayer graphene, the nomenclature is slightly different. A difference in the number of electrons
in one sublattice compared to the other corresponds to a layer-polarized state and not to a CDW because the
layer-polarized state does not
break the translational symmetry of the lattice. Among a plethora of superconducting orders in bilayer graphene,
the one that is obtained by pseudospin rotation from the layer-polarized state is the $s_{++}$ superconductor.
Following the argument in the previous paragraph, we conclude that the tendencies towards the
layer-polarized and $s_{++}$ superconducting orders are exactly the same
in bilayer graphene at half filling with an attractive Hubbard interaction.

\begin{figure}[ht]
\begin{center}
\includegraphics[width=0.49\textwidth]{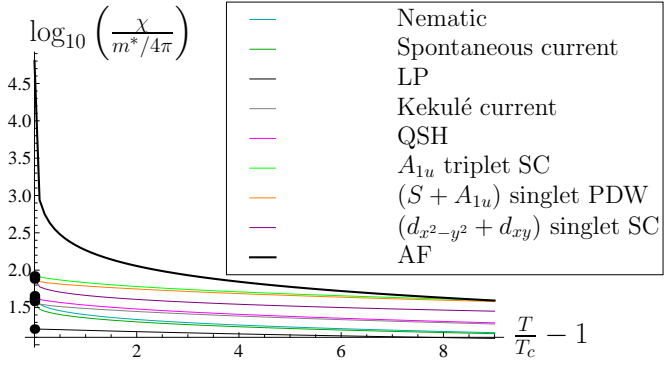}
\end{center}
  \caption{(Color online) Susceptibilities for the repulsive Hubbard model on the honeycomb bilayer.
    The AF susceptibility is the most dominant at all temperatures. The  dominance
    of the AF susceptibility over the others is a sign of growing AF correlations.
    Here, $\nu_3=0.178$,
    the bare couplings are $g_{A_{1g}}=g_{A_{2u}}=g_{E_\bK}=0.0560\times 4\pi/m^*$, and
    all others zero.  In this case, $t_c=0.01$.}\label{FigSusceptibilityAF}
\end{figure}

In addition, a repulsive Hubbard model is related to its attractive counter part with an equally strong
interaction. The mapping between the two is given by \cite {AuerbachBook}
\begin{eqnarray}
  c_{\uparrow} (\bR) &\to& \tilde c_{\uparrow} (\bR), \label {bup} \\
  c_{\downarrow} (\bR) &\to& (-1)^{\bR} \tilde c_{\downarrow}^\dagger (\bR). \label {bdown}
\end {eqnarray}
It is easy to demonstrate that, under these transformations, the kinetic term of a Hubbard
model on a bipartite lattice at half filling remains the same, while the interaction term changes sign.

The pseudospin symmetry, as well as the mapping of the repulsive Hubbard model to its attractive counterpart,
are also present in the honeycomb bilayer lattice. The question is then whether any of these properties survive in the
low-energy effective  field theory, in which the only degrees of freedom considered are those around
$\bK$ and $\bK'$. In the $\beta$ functions, Eq.\ \eqref {eq:gflow}, the pseudospin symmetry is not
apparent. Moreover, once we start from a set of bare interactions corresponding to the
Hubbard interaction, all nine couplings are generated. However,
as we will now demonstrate, both the pseudospin symmetry and the connection between the attractive and
repulsive Hubbard models are present in the RG flow. These manifest themselves through
certain linear combinations of the couplings that are invariant when the bare interactions correspond
to the Hubbard model.

We start by rewriting the mapping given by Eqs.\ \eqref {bup} and \eqref {bdown}
for our low-energy effective theory. Fields  with spin up transform as
\begin {eqnarray}
  \psi^{(b1/b2)}_{\pm \bK \uparrow} (\bq) \to \tilde \psi^{(b1/b2)}_{\pm \bK \uparrow} (\bq), \label {psiup}
\end {eqnarray}
while those with spin down transform according to
\begin {eqnarray}
  \psi^{(b1)}_{\pm \bK \downarrow} (\bq) &\to& \phantom {-} \tilde \psi^{(b1)*}_{\mp \bK \downarrow} (-\bq), \label {psidown1} \\
  \psi^{(b2)}_{\pm \bK \downarrow} (\bq) &\to& - \tilde \psi^{(b2)*}_{\mp \bK \downarrow} (-\bq). \label {psidown2}
\end {eqnarray}
This mapping leaves the kinetic term, Eq.\ \eqref {Hkin},
invariant, as it should, but it changes the interaction term, Eq.\ \eqref {Hint},
\begin {eqnarray}
  \frac 1 2 \sum_S g_S \left ( \psi^\dagger S \psi \right )^2 \to
    \frac 1 2 \sum_S \tilde g_S \left (\tilde \psi^\dagger S \tilde \psi \right )^2, \label {mapHint}
\end {eqnarray}
where each coupling constant $\tilde g_S$ is a linear combination of the coupling constants $g_S$ before the
transformation. We find that four of the nine coupling constants,
$g_{A_{2g}}$, $g_{E_g}$, $g_{A_{1u}}$, and $g_{A_{1\bK}}$,
do not change sign, i.e.,  $g_{i} \to \tilde g_{i}$, under this mapping. We
call these coupling constants ``even.'' In addition, there are two linear combinations,
\begin {eqnarray}
  g_{a} &=& g_{A_{1g}} + g_{A_{2u}} + \frac 1 2 g_{E_\bK}, \label {godda} \\
  g_{b} &=& g_{A_{1g}} - g_{A_{2u}} + g_{E_u} +  g_{A_{2\bK}}, \label {goddb}
\end {eqnarray}
which change sign under the mapping, i.e., $g_{a/b} \to - \tilde g_{a/b}$. We will refer to these
as ``odd.'' Finally, there are three remaining linearly independent combinations of the coupling constants,
\begin {eqnarray}
  \delta g_1 &=& g_{A_{1g}} - g_{A_{2u}} - 2 g_{E_u}, \label {deltag1} \\
  \delta g_2 &=& g_{A_{1g}} - g_{A_{2u}} - 2 g_{A_{2\bK}}, \label {deltag2} \\
  \delta g_3 &=& g_{A_{1g}} + g_{A_{2u}} - 4 g_{E_\bK}, \label {deltag3}
\end {eqnarray}
which are neither ``even'' nor ``odd,'' as they generate terms proportional to themselves as
well as terms proportional to ``even'' and ``odd'' coupling constants in the RG flow.
Clearly, a non-zero value of any $\delta g_i$ would spoil the
connection between the two models.

In the Hubbard limit, we notice that all three $\delta g_i$'s are zero at $\ell=0$. We
also see that, initially, all other couplings are exactly zero except for $g_a \sim 9U/4 \neq 0$. Therefore,
Eqs.\ \eqref {psiup}--\eqref {psidown2}, when applied to a repulsive Hubbard model
with interaction strength $U$, changes the sign of the
only non-zero coupling $\tilde g_a = - g_a$ and leaves all the other couplings zero.
This is precisely the bare interaction of an attractive Hubbard model with the same interaction strength.

So far, we have shown that the low-energy effective field theories for the repulsive and attractive Hubbard models in
bilayer graphene map onto each other, but only at the bare level. To show the equivalence at any $\ell$, we look
at the flow of the coupling constants, i.e., their linear combinations, Eqs.\ \eqref {godda}--\eqref {deltag3}. For the three
couplings, Eqs.\ \eqref {deltag1}--\eqref {deltag3}, we find that
\begin {eqnarray}
  \frac {\rmd \delta g_i}{\rmd \ell} = \beta_{\delta g_i} (\delta g_1, \delta g_2, \delta g_3)
    \stackrel {\delta g_j \to 0}{\longrightarrow} 0. \label {betadeltag}
\end {eqnarray}
The last arrow means that all three $\beta$ functions vanish when all three
$\delta g_i$'s are simultaneously zero. Since this is true at $\ell=0$ in the Hubbard limit, it
follows that no $\delta g_i$'s can be generated in the RG flow.

On the other hand, the ``even'' and ``odd'' coupling constants {\em do} flow under
RG, but there is a special structure to their $\beta$ functions. The four ``even'' couplings flow according to
\begin{eqnarray}
  \frac {\rmd g^{(e)}_i}{\rmd \ell} &=&  \sum_{a=1}^4 \bigg \lbrack \sum_{j,k=1}^4 g^{(e)}_j g^{(e)}_k \bar A^{(e/e)(a)}_{ijk} + \label {evengflowHubbard} \\
  && \qquad \quad  \sum_{j,k=1}^2 g^{(o)}_j g^{(o)}_k \bar A^{(o/o)(a)}_{ijk} \bigg \rbrack \Phi_{a}. \nonumber
\end{eqnarray}
The flows of the two ``odd'' couplings are given by
\begin{eqnarray}
  \frac {\rmd g^{(o)}_i}{\rmd \ell} &=& 2 \sum_{a=1}^4 \sum_{j=1}^2 \sum_{k=1}^4 g^{(o)}_j g^{(e)}_k \bar A^{(o/e)(a)}_{ijk}
    \Phi_{a}. \label {oddgflowHubbard}
\end{eqnarray}
The fact that all $\delta g_i$'s are zero for any $\ell$ has already
been incorporated,  and thus there are only six independent
couplings. The structure of Eqs.\ \eqref {evengflowHubbard} and
\eqref {oddgflowHubbard} makes them manifestly invariant under the
transformation,  \eqref {psiup}--\eqref {psidown2}. Therefore, an
RG flow obtained within the continuum low-energy effective field
theory corresponding to an attractive Hubbard model is described by
the same set of differential equations as its repulsive counterpart.
The flows of the ``even'' couplings are identical for the two cases,
while those for the ``odd'' couplings differ only by a sign. The
three couplings $\delta g_i$ are all zero at any $\ell$ for both
cases. Had the flow started from a point where at least one of the
$\delta g_i$'s were finite, this correspondence would have been
spoiled because additional terms appear in  Eqs.\ \eqref
{evengflowHubbard} and \eqref {oddgflowHubbard}.
\begin{figure}[ht]
  \centering
  \includegraphics[width=\columnwidth]{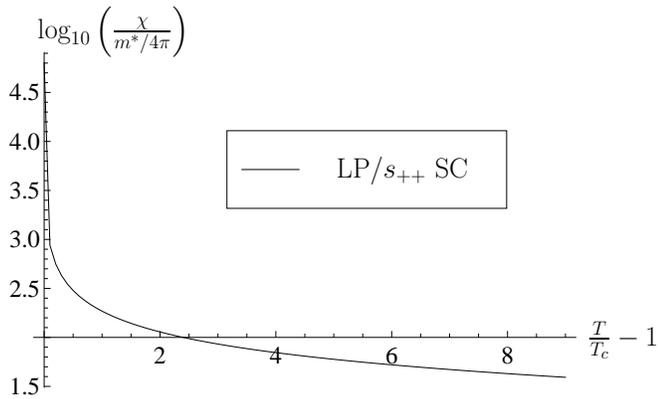}
  \caption{\label{FigsSusceptibilityLPSC1} Susceptibilities toward the layer-polarized
  and $s_{++}$ superconducting orders in the attractive Hubbard model.  In this case,
  both susceptibilities are equal.  The bare coupling constants used here are $g_{A_{1g}}=
  g_{A_{2u}}=2g_{E_\bK}=-0.0560\times 4\pi/m^*$, with all others zero.  The transition
  temperature in this case is $t_c=0.01$.}
\end{figure}

An immediate consequence of the mapping described here is that  physical quantities obtained in our
one-loop RG method
for a repulsive Hubbard model are related to those obtained from its attractive counterpart. For example,
the ``critical temperatures'' $t_c$ for the two models are the same.
Of course, the layer antiferromagnetic phase for  the repulsive Hubbard model and the
layer-polarized and $s_{++}$ superconducting phases
for the attractive Hubbard model both have the zero transition temperature because a continuous
$O (3)$ symmetry cannot be broken at any finite temperature in two dimensions.
The finite $t_c$ that we obtain within this approximate technique corresponds to
a gap scale that must be the same for both models.

Having demonstrated these special properties of the RG flow in the Hubbard limit, we now compare the
susceptibilities for the layer-polarized and $s_{++}$ superconducting states. The  $\alpha$ coefficients
for the corresponding source terms in the free energy, Eq.\ \eqref {eq:free energy},  are equal.
Therefore, it is
sufficient to look at the difference of the right hand sides of Eqs.\ \eqref {dlogDeltaph} and \eqref {dlogDeltapp}:
\begin {eqnarray}
  \frac {\rmd \log \Delta_{\text{LP}}}{\rmd \ell} - \frac {\rmd \log \Delta_{s_{++}}^{\text{SC}}}{\rmd \ell} \qquad \qquad \qquad
     \qquad \qquad \qquad \qquad  \label {DeltaLPminDeltaSC} \\
  = 2 \left ( g_{A_{1g}} -
    3 g_{A_{2u}} - 2 g_{E_u} -2 g_{A_{2\bK}} + 4 g_{E_\bK} \right ) (\Phi_1 + \Phi_4) \nonumber \\
  = 2 \left ( \delta g_1 + \delta g_2 - \delta g_3 \right ) (\Phi_1 + \Phi_4)
    \stackrel {\delta g_j \to 0}{\longrightarrow} 0. \nonumber
\end {eqnarray}
Since none of the $\delta g_i$'s are generated in the RG flow in the Hubbard limit, the source terms for the layer-polarized
and $s_{++}$ superconducting states flow in exactly the same way, their susceptibilities must diverge
with the same exponent. This proves that our one-loop RG treatment respects the pseudospin
symmetry of the Hubbard model at half filling. This argument remains valid for any value of the trigonal
warping, which does not break particle-hole symmetry, since no assumptions were made about $\Phi$ functions.

Notice that the Hubbard limit is not the only case in which the mapping and consequent pseudospin symmetry are realized. Any model in which the
bare values of all three $\delta g_i$'s are simultaneously zero will exhibit the above
correspondence as well. However, if we restrict ourselves to
microscopic density-density interaction Hamiltonians, in which case only three of the
four-fermion coupling constants are initially non-zero, the pseudospin symmetry is present only if
$g_{A_{1g}} = g_{A_{2u}} = 2 g_{E_{\bK}}$.
\begin{figure*}[ht]
  \centering
  \includegraphics[width=0.49\textwidth]{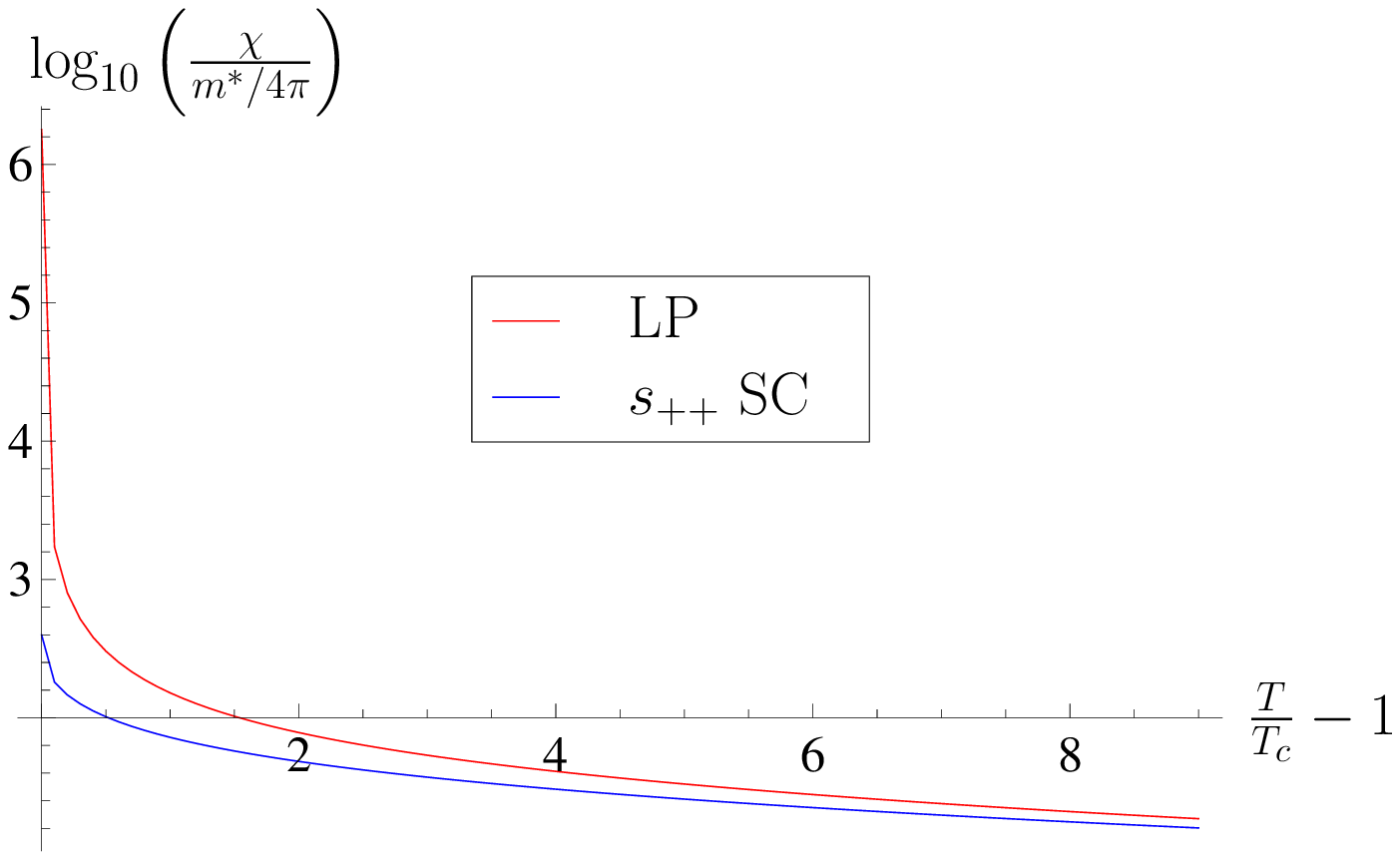}
  \includegraphics[width=0.49\textwidth]{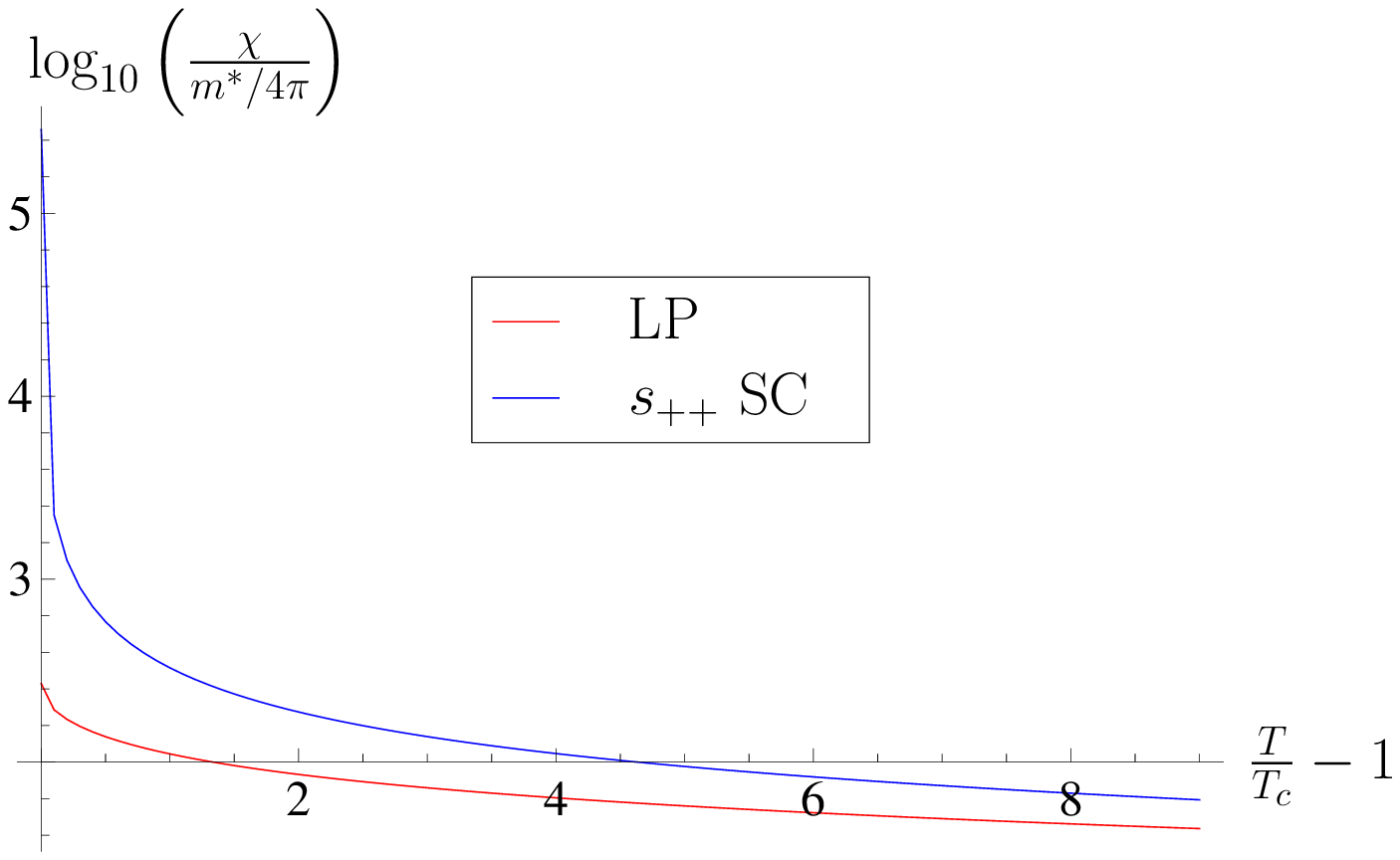}
  \caption{\label{FigsSusceptibilityLPSC2} (Color online) Comparison of the susceptibilities for
    the layer-polarized and $s_{++}$ superconducting orders
    in the attractive Hubbard model with an additional small $b_1$-$b_2$ interaction.
    In this case, $g_{A_{2u}}=\frac{1-\epsilon}{1+\epsilon}g_{A_{1g}}$, where $\epsilon=V/U$
    and $V$ is the microscopic $b_1$-$b_2$ interaction strength.  In both cases, the bare
    $g_{A_{1g}}=2g_{E_\bK}=-0.0560\times 4\pi/m^*$.
    (Left panel) When a small $b_1$-$b_2$ repulsion is added, the layer-polarized state is preferred.
    Its susceptibility diverges at $t_c$, while that of the superconducting state,
    as well as all other order parameters considered, reaches a finite value.
    Here, we take $\epsilon=0.1$, in which case $t_c=5.18\times 10^{-3}$.
    (Right panel) In the case of a small $b_1$-$b_2$ attraction, the
    opposite is true --- the susceptibility for the $s_{++}$ superconducting state diverges,
    while that for the layer-polarized state remains finite at $t_c$.
    Here, $\epsilon=-0.1$, in which case $t_c=0.0288$.}
\end{figure*}

At the end of this section we present numerically obtained
susceptibilities for various orders in the case when $g_{A_{1g}} =
g_{A_{2u}} = 2 g_{E_{\bK}}$. In Fig.\ \ref {FigSusceptibilityAF},
susceptibilities to various orders in the repulsive Hubbard model
are shown as functions of temperature. The AF susceptibility
dominates and is the only one to diverge at $T_c$. Reiterating what
was stated before, this divergence is an artifact of our one-loop RG
approximation. Nevertheless, one-loop RG correctly singles out the
state that is known to be the ground state at $T=0$. The $t_c$ that
we find should be thought of as a gap scale for the AF order. Figs.\
\ref {FigsSusceptibilityLPSC1} and \ref{FigsSusceptibilityLPSC2}
compare the layer-polarized and $s_{++}$ superconducting
susceptibilities.  In Fig.\ \ref{FigsSusceptibilityLPSC1}, an
attractive Hubbard model is studied. In Fig.\
\ref{FigsSusceptibilityLPSC2}, we consider the same model with an
additional small $b_1$-$b_2$ repulsion (left panel) or attraction
(right panel). This additional term violates the pseudospin symmetry
of the Hubbard Hamiltonian. In this case, the coupling constants are
the same as in the Hubbard model, except that now
$g_{A_{2u}}=\frac{1-\epsilon}{1+\epsilon}g_{A_{1g}}$, where
$\epsilon=V/U$ and $V$ is the microscopic $b_1$-$b_2$ interaction
strength. When this interaction is repulsive, the system favors the
layer-polarized state over the superconducting state analogous to,
for example, the findings of Ref.\ \onlinecite {VarmaPRL1988}.
Conversely, when this interaction is attractive, it favors
delocalization of the electron pairs and the concomitant
superconducting ground state. Our numerical results are in the
agreement this.

\subsection{Fixed ratios and broken symmetry phases} \label {SectionFixedRatios}
\begin{figure*}[ht]
\centering
\includegraphics[width=0.9\textwidth]{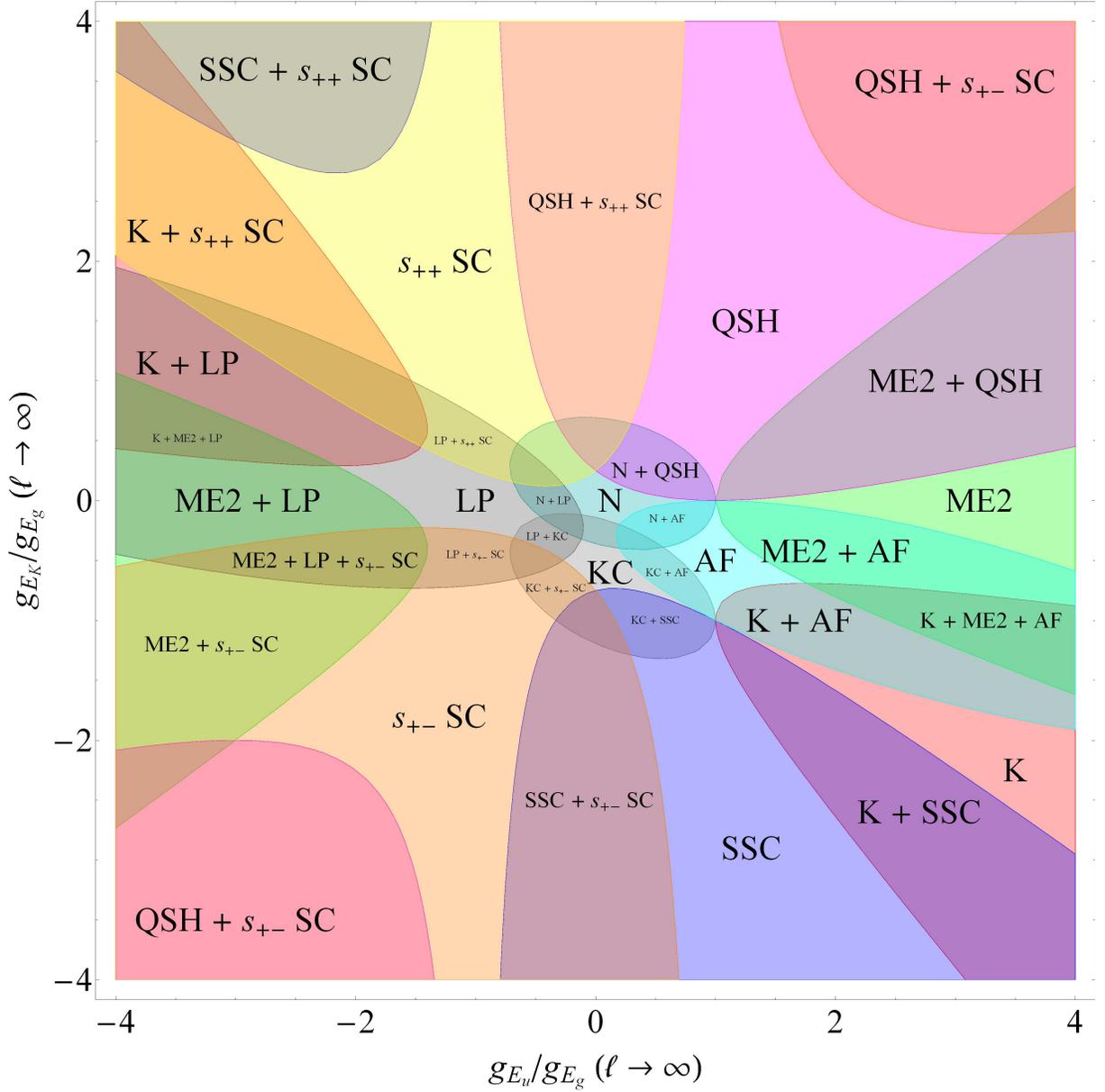}
\caption{\label{TargetPlane_PD}A plot of all of the phases found in
the fixed plane described by Equations
\eqref{FixedPlane_gA1g}-\eqref{FixedPlane_gA2K}. We find nematic (N,
$E_g$ charge), Kekul\'e (K, $A_{1K}$ charge), spontaneous current,
or magnetoelectric (ME2, $E_u$ charge), layer-polarized (LP,
$A_{2u}$ charge), Kekul\'e current (KC, $A_{2K}$ charge), staggered
spin current (SSC, $A_{1u}$ spin), antiferromagnetic (AF, $A_{2u}$
spin), quantum spin Hall (QSH, $A_{2g}$ spin), $s_{++}$
superconductor ($s_{++}$ SC, $A_{1g}$ singlet), and $s_{+-}$
superconductor ($s_{+-}$ SC, $A_{2u}$ singlet) states.  In addition
to this fixed plane, we also find four isolated fixed points, which
are described in the text.}
\end{figure*}

The list of phases found numerically in the previous section shows ordering trends for bilayer graphene only
for a certain kind of microscopic interaction that we believe is relevant in a realistic system. The question
is whether there are other possible ordered states in bilayer graphene when all 9 symmetry-allowed couplings
are included at $\ell=0$. One way to answer this question is to numerically explore the entire 9-dimensional space of
bare couplings for various trigonal warping parameters. Such an approach, although straightforward,
would require immense computational power and might even miss certain phases that are realized only
for specific bare interactions. Fortunately, there is another approach to the problem that we discuss in this section.
Instead of concentrating on the bare interactions, we look at what happens to the couplings and
susceptibilities at large $\ell$. This allows us to enumerate all the possible phases
regardless of the initial interactions.

Previously, we discussed the asymptotic behavior of the RG
equations at $t=t_c>0$. We know that at least one coupling will
diverge as $g_r (\ell) \sim e^{2 \ell}$. We divide all the other
couplings by that particular coupling and find the $\beta$ functions
for the ratios, $\rho_j^{(r)}=g_j/g_r$, to be
\begin {eqnarray}
  \frac {\rmd \rho_j^{(r)}}{\rmd \ell} &=& \frac {\dot g_j g_r - \dot g_r g_j}{g_r^2}= g_r (\ell) \sum_{k,l} \rho_k^{(r)} \rho_l^{(r)} \times   \label {betarho} \\
  &&  \qquad \sum_{a=1}^4
    \left ( A_{jkl}^{(a)} - A_{rkl}^{(a)} \rho_j^{(r)} \right ) \Phi_a (\nu_3 (\ell), t (\ell) ). \nonumber
\end {eqnarray}
Here, a dot over a coupling constant represents a derivative with respect to $\ell$.
In the large $\ell$ limit, these equations become
\begin{eqnarray}
  \dot \rho_j^{(r)} = \frac {8 t_c}{{\mathcal A}_{(r)}}  \sum_{k,l} \rho_k^{(r)} \rho_l^{(r)} \sum_{a=1}^2 \left ( A_{jkl}^{(a)} - A_{rkl}^{(a)} \rho_j \right ). \label {dotrho}
\end{eqnarray}
We now ask if these equations have any fixed points, or, in our terminology, ``fixed rays''. These are obtained
by demanding that the right hand sides of all 8 equations \eqref{dotrho} are simultaneously equal to zero. After finding the
fixed rays, we need to determine whether each ray is stable, unstable, or mixed by analyzing eigenvalues
of the stability matrix $S_{jk} = {\partial \dot \rho_j^{(r)}}/{\partial \rho_k^{(r)}}$. Since ${\mathcal A}_{(r)}$ is already defined
in Eq.\ \eqref {mathcalA} in terms of the ratios, the entire stability matrix has well-defined eigenvalues
for each ``fixed ray'' solution. In addition, the sign of ${\mathcal A}_{(r)}$ determines the sign of the
diverging coupling that we divide the others by; see Eq.\ \eqref {eq:g RGasymptotics}.

If we find that a ray is stable, then, if we start with the coupling constants sufficiently close to the fixed ray,
then the ratios of the couplings approach the given set of values as $\ell \to \infty$.  Such a flow leads to a
divergent susceptibility in at least one channel. If a ray
is mixed or unstable, then, in the absence of fine-tuning, the RG flow cannot take the couplings toward such a ratio; even if the flow starts in such a
direction for small $\ell$, it will be redirected toward some other ray that is stable.  We therefore conclude that all the solutions that have
even one positive eigenvalue in their stability matrix are physically irrelevant. It is possible that some rays are marginal
in certain directions, meaning that some of the eigenvalues of the stability matrix are
zero, and stable in others.  We do, in fact, find such physically relevant solutions.

Following the procedure described above for all possible choices of the divergent coupling, we find that the stable solutions of the
RG flow are situated either on a manifold that we call the ``target plane'' or on one of four isolated
fixed rays. The ``target plane'' represents a set of stable rays that are marginal in two directions
and stable in six others. The target plane and the phases corresponding to each point within
are shown in Fig.\ \ref {TargetPlane_PD}. We parametrize the plane in the following way. We choose as our parameters the following two coupling constant ratios:
\begin{eqnarray}
x&=&\lim_{\ell\rightarrow\infty}\frac{g_{E_u}}{g_{E_g}}\bigg|_{t=t_c}\\
y&=&\lim_{\ell\rightarrow\infty}\frac{g_{E_{\bK}}}{g_{E_g}}\bigg|_{t=t_c} \label{FixedPlane_gEK}.
\end{eqnarray}
Since, for certain fixed rays, $g_{E_u}$ and/or $g_{E_\bK}$ diverge,
while $g_{E_g}$ does not, these parameters take values in the
interval $(-\infty, \infty)$, including infinite values. With the
chosen parametrization, we express each coupling at large $\ell$ as
\begin {widetext}
\begin{eqnarray}
  \frac{g_{A_{1g}}}{G(\ell)}\bigg|_{t=t_c}= 0, \qquad g_{A_{2g}}\bigg|_{t=t_c}= \frac {(1+x+2y)^2}{C(x, y)} G(\ell),
    \qquad g_{E_{g}}\bigg|_{t=t_c}= \frac {-2 (1+x+2y)}{C(x,y)} G(\ell), \label{FixedPlane_gA1g} \\
  g_{A_{1u}}\bigg|_{t=t_c}= \frac {4 y^2}{C(x,y)} G(\ell), \qquad g_{A_{2u}}\bigg|_{t=t_c}= \frac {4 x}{C(x, y)} G(\ell),
    \qquad g_{E_{u}}\bigg|_{t=t_c}= \frac {-2 x (1+x+2y)}{C(x,y)} G(\ell), \\
  g_{A_{1\bK}}\bigg|_{t=t_c}= \frac {4 x y}{C(x,y)} G(\ell), \qquad g_{A_{2\bK}}\bigg|_{t=t_c}= \frac {4 y}{C(x,y)} G(\ell),
    \qquad g_{E_{\bK}}\bigg|_{t=t_c}= \frac {-2 y (1+x+2y)}{C(x,y)} G(\ell), \label{FixedPlane_gA2K}
\end{eqnarray}
\end {widetext}
where $C(x,y)$ is a square root of a quartic polynomial and the
``overall'' coupling $G(\ell) = \left \lbrack \sum_{j=1}^{9} g_j^2
\right \rbrack^{1/2}$ is a {\em positive definite} function of
$\ell$ that diverges as $\ell \to \infty$.  The expression for
$C(x,y)$ can be readily obtained from the definition of $G(\ell)$,
but is unwieldly, and thus we do not include it here. The ratios of
any two couplings at large $\ell$ depend only on $x$ and $y$,
although sometimes these ratios may be infinite.

In the special situation in which the
parameters $x$ and $y$ are infinite, but their ratio is finite, we may reparameterize $x$ and $y$ as
$x = R \cos \eta$ and $y = R \sin \eta$ and take the limit as $R \to \infty$. The only diverging couplings
in this case are $g_{A_{2g}}$, $g_{A_{1u}}$, $g_{E_u}$, $g_{A_{1\bK}}$, $g_{E_\bK}$. Note that,
for each $\eta$, we obtain the same stable ray at $\eta+\pi$. Due to the fact that any two opposite
points at infinity on the target plane are identical, we conclude that the target plane is
homeomorphic to a projective plane ${\mathbb R}P^2$. In  Fig.\ \ref {TargetPlane_PD}, some of the phases,
such as QSH, have hyperbolic phase boundaries and appear to exist in two disconnected parts of the phase
diagram. However, due to the fact that the opposite points in the target plane are identical, these
may be regarded as single and simply connected regions.

The values of $\rho_j^{(r)}=g_j/g_r$ are readily obtained from \eqref{FixedPlane_gA1g}-\eqref{FixedPlane_gA2K}. Without loss of generality we now set $g_r=g_{E_g}$ in Eq.\ \eqref{eq:g asymptotics}. We obtain
\begin{eqnarray}
\mathcal{A}_{(E_g)}=-6\frac{3+2x+3x^2+4y+4xy+8y^2}{1+x+2y}\frac{m^*}{4\pi}. \label{FixedPlane_A}
\end{eqnarray}
We may obtain $\mathcal{B}_{i,(E_g)}$ from Eqs. \eqref{eq:BCoeffPH} and \eqref{eq:BCoeffPP}. We can now determine the anomalous
dimensions of the symmetry-breaking source terms defined in
Eq.\ \eqref{eq:AnomDim}. Remarkably, we see that the
anomalous dimensions are continuous functions of the two parameters
$x$ and $y$. For each point in the target plane, we determine the
phases for which $\eta_i>1$, i.e., the inequality Eq.\ \eqref{eq: div susc} holds. If
more than one phase satisfies this inequality, then we list all
such phases regardless of the value of $\eta_i$.
As discussed before, whenever two or more susceptibilities diverge, we
cannot decide within our RG framework if the system chooses only one of these phases
or if there is a coexistence. The resulting list of phases is shown in Figure \ref{TargetPlane_PD}.
In addition to the phases we found earlier in Section \ref{Sec:DDInt}, namely the nematic (N), layer antiferromagnetic (AF),
quantum spin Hall (QSH), layer-polarized (LP), Kekul\'e current (KC), and
$s_{++}$ superconducting ($s_{++}$ SC) states, a few other phases are predicted as possible outcomes
of the RG flow if it ends on the target plane. These are:
{\it (a) Magnetoelectric phase (ME2)}: The order parameter for this phase transforms according to the $E_u$ charge
    representation.  In this phase, currents forming a bow-tie pattern within a plaquette appear. Like the nematic phase,
    this phase is gapless, but it reconstructs the low lying spectrum by lifting two of the four Dirac cones.
{\it (b) Kekul\'e state (K)}: In this phase, a supercell made of three unit cells is formed, much like the Kekul\'e current phase. The difference
    is that, in this phase, there are no currents.  Instead, there is a modification of the hopping integrals such that the hoppings in one unit cell are unchanged, while, in the two other unit cells, the hoppings on alternating bonds are changed\cite{HouPRL2007}. The phase is gapped.
{\it (c) Staggered spin current state (SSC)}: This phase is characterized by circulating spin currents in
    each plaquette flowing in opposite directions in each layer. This phase is not gapped, corresponds to a compensated semimetal, and the order parameter belongs to the $A_{1u}$ spin representation.
{\it (d) $s_{+-}$ superconducting state ($s_{+-}$ SC)}: Since a particle-particle susceptibility diverges in this case, a
    superconducting gap opens on both layers. The gaps are, however, not independent; they have opposite signs. The order parameter of this phase is a (charge 2) $A_{2u}$ spin singlet.

Strictly speaking, when either $x$ or $y$ becomes infinite or they satisfy $1+x+2y=0$, we are
not allowed to divide by $g_{E_g}$ as this coupling is not divergent.
It shows up in Eq.\ \eqref {FixedPlane_A} as a divergent ${\mathcal A}_{(E_g)}$.
Instead, these cases are explored by dividing by some other coupling. We follow the same procedure as described above in the case where we divided by $g_{E_g}$.  Interestingly, since both ${\mathcal A}_{(E_g)}$
and ${\mathcal B}_{i,(E_g)}$ diverge in the same way, the $\eta_i$'s are independent of the
choice of the coupling that we divide by.

In addition to the target plane, we also find the following four isolated stable fixed points.
\begin {itemize}
  \item [$R_1$:]
    \begin{eqnarray}
      \lim_{\ell\rightarrow\infty}\frac{g_{A_{1g}}}{g_{E_g}}\bigg|_{t=t_c}&=&3,\nonumber\\
      \lim_{\ell\rightarrow\infty}\frac{g_{j}}{g_{E_g}}\bigg|_{t=t_c}&=&1\;\;\forall\;j\neq A_{1g},
    \end{eqnarray}
    with $g_{E_g}(\ell\to\infty)>0$.  In this case, only the ferromagnetic ($A_{1g}$ spin) susceptibility
    diverges.
  \item [$R_2$:]
    \begin{eqnarray}
      \lim_{\ell\rightarrow\infty}\frac{g_{j}}{g_{A_{2g}}}\bigg|_{t=t_c}&=&0\;\;\forall\;j\neq A_{2g},
    \end{eqnarray}
    and $g_{A_{2g}}(\ell\to\infty)<0$. The only divergent susceptibility in this case is towards the anomalous quantum Hall state\cite{HaldanePRL1988} ($A_{2g}$ charge).  Here, charge currents circulate in each layer\cite{NandkishorePRB2010}, and in the same direction in both layers.
  \item [$R_3$:]
    \begin{eqnarray}
      \lim_{\ell\rightarrow\infty}\frac{g_{j}}{g_{A_{1u}}}\bigg|_{t=t_c}&=&0\;\;\forall\;j\neq A_{1u},
    \end{eqnarray}
    and $g_{A_{1u}}(\ell\to\infty)<0$. This yields a loop current order\cite{ZhuArXiv}, or ``orbital antiferromagnet'' ($A_{1u}$ charge). Like the above phase, there are charge currents circulating in each layer, but in opposite directions.  Note that the order parameter, $\tau_3 1$, can be thought of as a chemical potential shift with opposite signs in each valley.  Therefore, at weak coupling, this phase corresponds to a compensated semimetal with electron and hole pockets.
  \item [$R_4$:]
    \begin{eqnarray}
      \lim_{\ell\rightarrow\infty}\frac{g_{j}}{g_{A_{1g}}}\bigg|_{t=t_c}&=&0\;\;\forall\;j\neq A_{1g},
    \end{eqnarray}
    with $g_{A_{1g}}(\ell\to\infty)<0$. Although we would intuitively
    expect this fixed point to favor a superconducting state, we find no particle-particle susceptibilites diverging. Only the $A_{1g}$ charge susceptibility, or equivalently the electronic compressibility, diverges.  Therefore, we conclude that the system enters a phase segregated state.
\end {itemize}

We can now make a connection between the results obtained in previous sections and the analytic
results obtained here. For the set of initial couplings and parameters analyzed in Secs.\ \ref{SecFwdNematic}--\ref{Sec:Hubbard}, the flow at $t_c$ always converges to the target
plane, and never to any of the isolated points $R_j$. In the case of forward scattering only at $\ell=0$, and in the absence of trigonal warping, none of the couplings
from the $u$ and $\bK$ representations are generated. The flow always ends at the point, $(x, y) = (0, 0)$, in the target
plane, which corresponds to a pure nematic state. With trigonal warping included, $u$ representation couplings are
generated, even when we start with forward scattering only. However, we still do not generate any of the $\bK$ representation couplings. This means that the end point of the flow at $t_c$ is restricted to
the $y=0$ line in the target plane. Decreasing the initial coupling strength $g_{A_{1g}}$, while holding the bare $v_3$ fixed, causes $t_c$ to decrease. At the same time, $x$ increases. We always find that $x<1$. As seen from Fig. \ref{TargetPlane_PD}, these points correspond to a pure nematic order. However, as $t_c$ is lowered, the point in the target plane moves closer to $x=1$, which is the intersection of the AF and QSH regions in the target plane. Upon reaching $t_c=0$ exactly, we find that the nematic order
parameter is absent while the AF and QSH susceptibilities become divergent. This is illustrated in Fig. \ref{fig:Crit v3 for gA1g}(b). Note that there is no point at which the only
two diverging susceptibilities are AF and QSH, either in the target plane or as one of the four isolated
points.  This is because the asymptotic behavior of the $\Phi$ functions is different at $t=0$ and our analysis, in which we assumed that $t_c>0$, does not apply there.

When the RG flow begins with a finite backscattering, i.e., $g_{E_\bK}\neq 0$, all 9 couplings are generated. We find that our numerical results correspond to points in the target plane where $y \neq 0$. With the physical constraints we impose on the initial couplings, $|g_{A_{2u}}| \le g_{A_{1g}}$, $|g_{E_\bK}| \le g_{A_{1g}}/2$ and $g_{A_{1g}}$ positive, only the central region of the target plane is approached. With these constraints, we do not find any set of initial couplings for which K, SSC, or ME2 phases appear.

In the previous sections, the flows never reach any of the isolated points $R_j$. However, one can
expect that the flow will tend to one of these points if one starts with bare couplings sufficiently close to
the associated ray and with a large initial value of $\ell$.
To confirm this, we analyzed the flow equations with no trigonal warping and $g_{A_{1g}}(\ell=0)<0$.
For sufficiently large initial values of the interaction $\frac{m^*}{4\pi}g_{A_{1g}}(\ell =0) < \frac{m^*}{4\pi}g_{A_{1g}}^c \approx -0.13$, the flow takes the couplings
towards the $R_4$ fixed ray. As stated before, this represents a compressibility instability. However, when $g_{A_{1g}} (\ell =0) > g_{A_{1g}}^c$ the couplings diverge toward the nematic fixed ratio, i.e., our flows end up on the target plane.

The symmetry properties of the Hubbard limit have consequences for the asymptotic behavior
of the RG flows studied in this section. When all three couplings, $\delta g_i$, Eqs.\ \eqref {deltag1}--\eqref {deltag3} are absent at the bare level, we have shown that they remain zero throughout the entire flow. As argued above, the ratios of couplings at $t_c$ must lie either on the target plane or at one of the four isolated points $R_j$.
In the target plane, the condition, $\delta g_i=0$, is satisfied only when $x = -2 y$. This defines a line of fixed points
(strictly speaking a circle, since two points at infinity are equivalent). As shown in Sec.\ \ref{Sec:Hubbard}, because $\delta g_i=0$, the susceptibilities
towards the layer-polarized and $s_{++}$ superconducting states are identical.  Therefore, this holds along the entire line $x=-2y$.
For the repulsive Hubbard model, a wide range of initial conditions, $10^{-8}<t_c<1$, maps onto the segment of this fixed line that lies within the AF-only region. These results were also used in studying the attractive Hubbard model due to the $U\rightarrow -U$ correspondence presented in Sec.\ \ref{Sec:Hubbard}. The only
difference is that both $x$ and $y$ change their sign under this mapping. The resulting fixed
points are therefore situated in the region where the layer-polarized and $s_{++}$ superconducting
orders overlap.

In addition to the fixed line that is part of the target plane, the
condition that all three $\delta g_i$'s are zero is satisfied at the
isolated fixed point $R_1$. However, we never find a flow toward
that point for any set of bare couplings studied here.

\section{Effect of a perpendicular electric field on the phase boundaries}
\begin{figure}[t]
\centering
\includegraphics[width=.9\columnwidth]{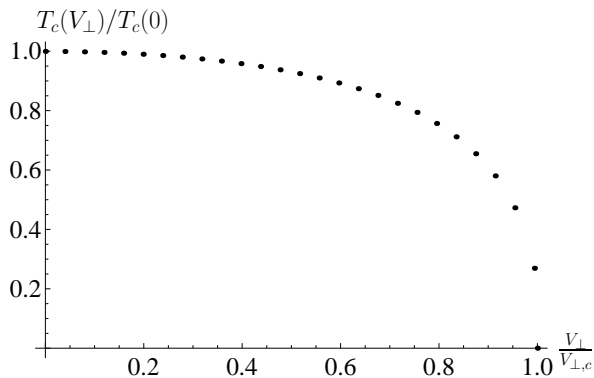}
\caption{\label{Tc_vs_V}Plot of the critical
temperature $T_c$ as a function of the layer energy
difference $V_\perp$, under such conditions that $T_c(0)=0.01\Lambda^2/2m^*$.
The critical temperature becomes zero when $V_\perp=V_{\perp,c}=4.93\times
10^{-3}\Lambda^2/2m^*$, which corresponds to an applied electric field of
$\sim 16\text{ mV/nm}$.}
\end{figure}

We now consider the effect that applying a perpendicular electric
field has on the phase boundaries of our system.  This field creates
an energy difference between the two layers of the sample, thus
introducing a new term into the Hamiltonian,
\begin{equation}
\mathcal{H}_{E_\perp}=V_\perp\sum_{|\bk|<\Lambda}\sum_{\sigma=\uparrow,\downarrow}
\psi^{\dag}_{\bk\sigma}1\sigma_3\psi_{\bk\sigma}.
\end{equation}
We state the effects that this has on the Green's function and on the
associated identities that we use in Appendix \ref{App:GF_PerpEField}, and simply quote the main
results here.  The RG flow equations for the coupling constants
become
\begin{equation}
\frac{dg_i}{d\ell}=\sum_{j,k}g_j g_k\sum_{a=1}^6 A_{ijk}^{(a)}\Phi_a(\nu_3(\ell),v(\ell),t(\ell)), \label{eq:gflow_ef}
\end{equation}
where, in addition to the dimensionless parameters for the $\Phi$ functions that were defined before, we have one new parameter,
\begin{equation}
v(\ell)=\frac{V_\perp(\ell)}{\Lambda^2/2\ms}.
\end{equation}
The $\Phi$ functions are given by Eqs. \eqref{Phi_EField_1}--\eqref{Eq:Qpm_EField}.

In addition, the energy difference $V_\perp$ has a nontrivial behavior under rescaling; it obeys the flow equation,
\begin{equation}
\frac{dv}{d\ell}=2v\left [1+F(\nu_3,v,t)\sum_{i}b_i g_i\right ], \label{Eq:VFlow}
\end{equation}
where the coefficients $b_i$ are given in Appendix C, and the function $F$ is given by Eq. \eqref{Eq:FFunc}.

We studied the behavior of the critical
temperature as a function of $v$ in a case where we know that the
system enters the nematic phase when $v=0$, namely when all coupling
constants are zero except for $g_{A_{1g}}>0$.  We assume that the
energy scale $\frac{\Lambda^2}{2\ms}=200\text{ meV}$, and that the
critical temperature at zero field is $T_c=2\text{ meV}$; i.e.
$t_c=0.01$.  To determine $t_c$ for a given initial $g_{A_{1g}}$, we
start with a high value of $t$ and integrate our RG flow equations
numerically up to a large value of $\ell$, say 10.  If we do not
encounter any divergences in the flows, we lower $t$ and integrate
again.  We continue until we find the highest temperature at which
we encounter a divergence.  Under the stated conditions, the
behavior of $t_c$ as a function of $v$ is as shown in Fig.\
\ref{Tc_vs_V}.

We find that, at $v=4.93\times 10^{-3}$, $t_c$ becomes zero.
Converting this into an electric field using the stated energy scale
and the formula relating the applied electric field to the size of
the gap in the spectrum, $\Delta=\frac{dE}{k}$, where $d$ is the
distance between the two graphene layers and $k\approx 3$ is a
factor accounting for imperfect screening\cite{YacobyScience2010, YacobyPRL2010},
we find that the electric field required to drive $t_c$ to zero is
$\sim 16\text{ mV/nm}$.

We have also determined which phase the system enters at all points on
the curve in Fig. \ref{Tc_vs_V} at which $T_c\neq 0$.  We did
this by deriving the RG flow equations for the source terms and the
formula for the free energy per unit area using the same procedures
as before, but with properly modified Green's functions, whose forms
are given in Appendix B.  We then numerically integrate the RG
flow equations, and, from these solutions, determine the
susceptibilities to the phases corresponding to each source term
just above the critical temperature; the phase with the highest
susceptibility is considered to be the phase that is present.  Using
this procedure, we determined that, for all $V_\perp<V_{\perp,c}$, the
system enters the nematic phase.

\section{Discussion}

The key result of this work is the identification of the conditions
on the electron-electron interactions under which various electronic
ordering tendencies, if any, dominate in half-filled bilayer
graphene. Our results for the ordered states are summarized in
Figs.\ \ref{FigPhaseDiagramTv3} and \ref{TargetPlane_PD}. Aside from
our use of one-loop RG, no further approximations are made.
Therefore, our results can be stated rigorously at the level of
mathematical theorems. While a large number of phases is, in
principle, possible in the entire 9-dimensional space of couplings,
as one can see from Fig. \ref{TargetPlane_PD}, our assertion is that
the electronic nematic appears to be the unique dominant instability
when forward scattering dominates. Similarly, the layer
antiferromagnet appears upon inclusion of sufficiently strong back
and layer imbalance scattering.

A similar approach in one spatial dimension\cite{DzyaloshinskiiLarkin1972} results in divergences in
the scattering amplitudes at finite temperature, naively suggesting
a finite temperature phase transition, which we know cannot happen.
Nevertheless, among many possibilities, the method does identify the
correct channels for which long, but finite, correlation lengths develop.
For example, the low-energy effective field theory for the
course-grained half-filled Hubbard model does correctly determine
that the dominant correlations appear in either the pairing
(attractive $U$) or AF (repulsive $U$) channel\cite{SchultzPRL1990}. Away from any special
filling, a metallic state is also correctly predicted \cite {BychkovGorkovDzyaloshinskii1966}.

We view our RG results for the half-filled Hubbard model similarly.
While the method correctly determines the dominant ordering
tendency, there can be no finite temperature phase transition in 2D
to either the AF state or the SC/LP state. A continuous spin
SU(2) symmetry in the former case, or a continuous pseudospin
symmetry in the latter case,  would have to be broken at finite temperature, which
we know cannot happen. Therefore, if the RG procedure had been performed exactly,
none of these susceptibilities would have diverged as long as the
temperature was finite. Interestingly, in this regard, the nematic
state is different. This is because, when trigonal warping is included, as it
is in our model, the broken rotational symmetry is
{\em discrete} and thus it is possible to have a finite-temperature transition
into this phase in 2D.
As argued previously\cite{VafekYangPRB2010}, this transition is
continuous and belongs to the 3-state Potts model universality class\cite{WuRMP1982}.
Nevertheless, the {\it non-mean-field} exponents determined
approximately from our fermionic model at one loop should not be
expected to be accurate. It would be very interesting to see whether
going to higher order either improves the accuracy of the exponents in the case of the
nematic state or eliminates the finite-temperature phase transition
altogether for the case of $O(3)$ order parameters. The effects of
(weak) disorder have  not been addressed here either. These
considerations may be important in fully understanding the current
experimental results\cite{YacobyScience2010, YacobyPRL2010,Mayorov2011,VelascoNatNano2012,
FreitagPRL2012,VeliguraPRB2012,BaoArXiv}.

Even if the RG method used here is not without its limitations, it
is unbiased and capable of systematically treating the leading
instabilities in both particle-hole and pairing channels. In
fact, for a large range of temperatures above the transition
temperature, the couplings saturate to small finite values as all
modes are eliminated, giving full justification to our method. In
the special case in which our continuum field theory corresponds to
the weak-coupling  honeycomb bilayer Hubbard
model, we recover some of its  non-trivial, exactly known,
properties. This gives further support for the validity of our
results.

\acknowledgements
This work was supported by the NSF CAREER award under Grant No.
DMR-0955561,  NSF Cooperative Agreement No. DMR-0654118, and
the State of Florida. The authors would also like to thank the
KITP-UCSB Research Program, ``The Physics of Graphene,'' where
part of this work was completed, for hospitality. The work at
KITP was supported in part by NSF Grant No. PHY-0551164.

\appendix
\begin{widetext}
\section{Asymptotic behavior of the $\Phi$ functions} \label{App:AsympPhi}
The finite-temperature Green's function that is used to calculate
the RG flows for infinitesimal symmetry breaking source terms is
\begin{eqnarray}
G_{\bk}(i\omega_n)&=&\left(-i\omega_n1_8+\frac{1}{2m^*}(k^2_x-k^2_y)1\sigma_1 1+v_3k_x\tau_3\sigma_1 1+
\frac{1}{m^*}k_xk_y\tau_3\sigma_2 1-v_3k_y1\sigma_2 1\right)^{-1}\\
&=&\frac{1}{2}\sum_{s=\pm}\left(1+s\tau_3\right)
\frac{i\omega_n1+(\frac{1}{2m*}k^2\cos2\theta_{\bk}+sv_3k\cos\theta_{\bk})\sigma_1+(s\frac{1}{2m^*}k^2\sin2\theta_{\bk}-v_3k\sin\theta_{\bk})\sigma_2}{\omega^2_n+
\frac{1}{4{m^*}^2}k^4+v^2_3k^2+s\frac{1}{m^*}v_3k^3\cos3\theta_{\bk}}1.
\end{eqnarray}
Throughout the Appendix, we will use the notation, $\tau_i\sigma_j s_k$, for the ($8\times 8$) matrices that appear in our expressions; the Pauli matrices operate in valley, layer, and spin space, respectively.  We find the following identity useful when calculating the flow
equations:
\begin{eqnarray}
&&\int_{\Lambda(1-d\ell)}^{\Lambda}\frac{dk
k}{2\pi}\frac{1}{\beta}\sum_{n=-\infty}^{\infty}\int_{-\pi}^{\pi}\frac{d\theta_{\bk}}{2\pi}G_{\bk}(i\omega_n)\otimes
G_{\pm\bk}(\pm i\omega_n)=\nonumber\\
&&d\ell\frac{m^*}{8\pi}  \left[\mp1_8\otimes
1_8\left[\Phi_{1}\left(\nu_3,t\right)+\Phi_{2}\left(\nu_3,t\right)\right]
+\frac{1}{2}\left(1\sigma_1 1\otimes 1\sigma_1 1+\tau_3\sigma_2 1\otimes
\tau_3\sigma_2 1\right)\left[\Phi_{3}\left(\nu_3,t\right)+\Phi_{4}\left(\nu_3,t\right)\right]
\right]\nonumber\\
&+&d\ell\frac{m^*}{8\pi}\left[-\tau_31_4\otimes
\tau_31_4\left[\Phi_{1}\left(\nu_3,t\right)-\Phi_{2}\left(\nu_3,t\right)\right]
\pm\frac{1}{2}\left(\tau_3\sigma_1 1\otimes
\tau_3\sigma_1 1+1\sigma_2 1\otimes1\sigma_2 1\right)\left[\Phi_{4}\left(\nu_3,t\right)-\Phi_{3}\left(\nu_3,t\right)\right]
\right]. \label{2GF_Identity}
\end{eqnarray}
The $\Phi$ functions are defined as
\begin{eqnarray}\label{eq:Phi functions definition}
\Phi_1(\nu_3,t)&=&\frac{1}{2\pi}\frac{1}{t}\int_{-1}^{1}\frac{dx}{\sqrt{1-x^2}}\Upsilon_1(x,\nu_3,t),
\label{Eq:Phi1} \\
\Phi_2(\nu_3,t)&=&\frac{1}{\pi}\frac{1}{\nu_3}\int_0^1\frac{dx}{\sqrt{1-x^2}}\frac{1}{x}
\Upsilon_2(x,\nu_3,t), \\
\Phi_3(\nu_3,t)&=&
\frac{1}{\pi}\frac{1-\nu_3^2}{\nu_3}\int_0^1\frac{dx}{\sqrt{1-x^2}}\frac{1}{x}
\Upsilon_3(x,\nu_3,t), \\
\Phi_4(\nu_3,t)&=&\frac{1}{2\pi}\frac{1}{t}\int_{-1}^{1}\frac{dx}{\sqrt{1-x^2}}
\Upsilon_4(x,\nu_3,t),
\end{eqnarray}
where
\begin{eqnarray}
\Upsilon_1(x,\nu_3,t) &=&\frac{1}{\cosh^2\left(\frac{Q_+}{2t}\right)}+\frac{2t}{Q_+}\tanh\left(\frac{Q_+}{2t}\right), \\
\Upsilon_2(x,\nu_3,t) &=&
\sum_{\lambda=\pm}\lambda Q_{\lambda}\tanh\left(\frac{Q_{\lambda}}{2t}\right), \\
\Upsilon_3(x,\nu_3,t)
&=&-\sum_{\lambda=\pm}\frac{\lambda}{Q_{\lambda}}\tanh\left(\frac{Q_{\lambda}}{2t}\right), \\
\Upsilon_4(x,\nu_3,t)
&=&\frac{-1}{\cosh^2\left(\frac{Q_+}{2t}\right)}+\frac{2t}{Q_+}\tanh\left(\frac{Q_+}{2t}\right), \label{Eq:Upsilon4}
\end{eqnarray}
and
\begin{eqnarray}
Q_{\pm}&=&\sqrt{1+\nu^2_3\pm 2x\nu_3}. \label{Eq:Qpm}
\end{eqnarray}

In the limit of $v_3=0$ we have
\begin{eqnarray}
\Phi_1(0,t)&=&\Phi_2(0,t)=\tanh\frac{1}{2t}+\frac{1}{2t}\frac{1}{\cosh^2\frac{1}{2t}}\\
\Phi_3(0,t)&=&\Phi_4(0,t)=\tanh\frac{1}{2t}-\frac{1}{2t}\frac{1}{\cosh^2\frac{1}{2t}}.
\end{eqnarray}
In the limit of $T=0$ we have
\begin{eqnarray}
\Phi_1(\nu_3,0)&=&\Phi_4(\nu_3,0)=\frac{2}{\pi}\frac{1}{1+\nu_3}K\left(\frac{4\nu_3}{(1+\nu_3)^2}\right)\\
\Phi_2(\nu_3,0)&=&\frac{2}{\pi}\frac{1}{\nu_3}\frac{(1-\nu_3)^2}{1+\nu_3}
\left(\Pi\left(\frac{2\nu_3}{1+\nu^2_{3}},\frac{4\nu_3}{(1+\nu_3)^2}\right)-
K\left(\frac{4\nu_3}{(1+\nu_3)^2}\right)\right)
\\
\Phi_3(\nu_3,0)&=&\frac{2}{\pi}\frac{1-\nu_3}{\nu_3}\left(K\left(\frac{4\nu_3}{(1+\nu_3)^2}\right)-
\frac{(1-\nu_3)^2}{1+\nu^2_{3}}\Pi\left(\frac{2\nu_3}{1+\nu_3^2},\frac{4\nu_3}{(1+\nu_3)^2}\right)\right).
\end{eqnarray}
Here, the complete elliptic integrals of the first, $K(x)$, and the
third, $\Pi(x,y)$, kind are defined as
\begin{eqnarray}
K(x)&=&\int_0^{\frac{\pi}{2}}\frac{d\phi}{\sqrt{1-x\sin^2\phi}}\\
\Pi(x,y)&=&\int_0^{\frac{\pi}{2}}\frac{d\phi}{(1-x\sin^2\phi)\sqrt{1-y\sin^2\phi}}.
\end{eqnarray}
Physically, the logarithmic singularity associated with $K(x)$ has
its origin in the logarithmic divergence of the density of states at
the van Hove point, where the lines of constant energy near each
$\bK$ point change from single to four closed contours. These log
singularities appear only at $t=0$, for $t>0$ they are smeared out.
Because the divergences are integrable, they aren't the cause of
divergence of $g_i's$ in (\ref{eq:gflow}). Instead, the coupling
constants receive a ``boost'' at $\ell$ where $\nu_3(\ell)=1$.

\section{Green's functions in the presence of an applied perpendicular electric field} \label{App:GF_PerpEField}
In the presence of an applied electric field, the Green's function becomes
\begin{eqnarray}
G_{\bk}(i\omega_n)=\left
(-i\omega_n1_8+d^x_{\bk}1\sigma_11+v_3k_x\tau_3\sigma_11+d^{y}_{\bk}\tau_3\sigma_21-v_3k_y1\sigma_21+V_\perp
1\sigma_31\right )^{-1} \cr =\frac{1}{2}\sum_{s=\pm}\left(
1+s\tau_3\right )
\frac{i\omega_n1+(\frac{1}{2m*}k^2\cos{2\theta}+
sv_3k\cos{\theta})\sigma_1+(s\frac{1}{2m^*}k^2\sin{2\theta}-v_3k\sin{\theta})\sigma_2+V_\perp
1\sigma_3}{\omega^2_n+
\frac{1}{4{m^*}^2}k^4+v^2_3k^2+s\frac{1}{m^*}v_3k^3\cos{3\theta}+V_\perp^2}1.
\end{eqnarray}
The generalization of Equation \eqref{2GF_Identity} for this case is
\begin{eqnarray}
\int_{\Lambda(1-d\ell)}^{\Lambda}\frac{k\,dk}{2\pi}\frac{1}{\beta}\sum_{n=-\infty}^{\infty}\int_{0}^{2\pi}\frac{d\theta_{\bk}}{2\pi}\,G_{\bk}(i\omega_n)\otimes G_{\pm\bk}(\pm i\omega_n)= \cr
\frac{\ms}{8\pi}\,d\ell [(\mp 1_8\otimes 1_8-\tau_3 1_4\otimes\tau_3 1_4)\Phi_1(\nu_3,v,t)+(\mp 1_8\otimes 1_8+\tau_3 1_4\otimes\tau_3 1_4)\Phi_2(\nu_3,v,t) \cr
+\tfrac{1}{2}(1\sigma_11\otimes 1\sigma_11\mp\tau_3\sigma_11\otimes\tau_3\sigma_11\mp 1\sigma_21\otimes 1\sigma_21+\tau_3\sigma_21\otimes\tau_3\sigma_21)\Phi_3(\nu_3,v,t) \cr
+\tfrac{1}{2}(1\sigma_11\otimes 1\sigma_11\pm\tau_3\sigma_11\otimes\tau_3\sigma_11\pm 1\sigma_21\otimes 1\sigma_21+\tau_3\sigma_21\otimes\tau_3\sigma_21)\Phi_4(\nu_3,v,t) \cr
+(1\sigma_31\otimes 1\sigma_31+\tau_3\sigma_31\otimes\tau_3\sigma_31)\Phi_5(\nu_3,v,t) \cr
+(1\sigma_31\otimes 1\sigma_31-\tau_3\sigma_31\otimes\tau_3\sigma_31)\Phi_6(\nu_3,v,t)], \label{2GF_EField_Identity}
\end{eqnarray}
where the $\Phi$ functions are
\begin{eqnarray}
\Phi_1(\nu_3,v,t)&=&\frac{1}{2\pi}\frac{1}{t}\int_{-1}^{1}\frac{dx}{\sqrt{1-x^2}}\Upsilon_1(x,\nu_3,v,t), \label{Phi_EField_1} \\
\Phi_2(\nu_3,v,t)&=&\frac{1}{\pi}\frac{1}{\nu_3}\int_{0}^{1}\frac{dx}{\sqrt{1-x^2}}\frac{1}{x}\Upsilon_2(x,\nu_3,v,t), \\
\Phi_3(\nu_3,v,t)&=&\frac{1}{\pi}\frac{1-\nu_3^2}{\nu_3}\int_{0}^{1}\frac{dx}{\sqrt{1-x^2}}\frac{1}{x}\Upsilon_3(x,\nu_3,v,t), \nonumber\\ \\
\Phi_4(\nu_3,v,t)&=&\frac{1}{2\pi}\frac{1}{t}\int_{-1}^{1}\frac{dx}{\sqrt{1-x^2}}\Upsilon_4(x,\nu_3,v,t), \\
\Phi_5(\nu_3,v,t)&=&\frac{1}{2\pi}\frac{v^2}{t}\int_{-1}^{1}\frac{dx}{\sqrt{1-x^2}}\Upsilon_5(x,\nu_3,v,t), \\
\Phi_6(\nu_3,v,t)&=&\frac{v^2}{1-\nu_3^2}\Phi_3(\nu_3,v,t). \label{Phi_EField_6}
\end{eqnarray}
The $\Upsilon$ functions are
\begin{eqnarray}
\Upsilon_1(x,\nu_3,v,t)&=&\frac{2t}{Q_+}\tanh\left (\frac{Q_+}{2t}\right )+\frac{1}{\cosh^2\left (\frac{Q_+}{2t}\right )}, \\
\Upsilon_2(x,\nu_3,v,t)&=&\sum_{\lambda=\pm}\lambda Q_\lambda\tanh\left (\frac{Q_\lambda}{2t}\right ), \\
\Upsilon_3(x,\nu_3,v,t)&=&-\sum_{\lambda=\pm}\frac{\lambda}{Q_\lambda}\tanh\left (\frac{Q_\lambda}{2t}\right ), \\
\Upsilon_4(x,\nu_3,v,t)&=&\left (\frac{Q_+^{(0)}}{Q_+}\right )^2\left [\frac{2t}{Q_+}\tanh\left (\frac{Q_+}{2t}\right )-\frac{1}{\cosh^2\left (\frac{Q_+}{2t}\right )}\right ], \nonumber \\ \\
\Upsilon_5(x,\nu_3,v,t)&=&\frac{1}{(Q_+^{(0)})^2}\Upsilon_4(x,\nu_3,v,t),
\end{eqnarray}
where
\begin{eqnarray}
Q_\pm&=&\sqrt{1+\nu_3^2+v^2\pm 2x\nu_3}, \\
Q_\pm^{(0)}&=&\sqrt{1+\nu_3^2\pm 2x\nu_3}. \label{Eq:Qpm_EField}
\end{eqnarray}

One other identity that we will find useful is
\begin{eqnarray}
\int_{\Lambda(1-d\ell)}^{\Lambda}\frac{k\,dk}{2\pi}\frac{1}{\beta}\sum_{n=-\infty}^{\infty}\int_{0}^{2\pi}\frac{d\theta_{\bk}}{2\pi}\,G_{\bk}(i\omega_n)=\frac{\ms V_\perp}{2\pi}1\sigma_31 F(\nu_3,v,t)\,d\ell, \label{1GF_Identity}
\end{eqnarray}
where $F(\nu_3,v,t)$ is
\begin{equation}
F(\nu_3,v,t)=\frac{1}{\pi}\int_{-1}^{1}\frac{dx}{\sqrt{1-x^2}}\frac{1}{Q_+}\tanh\left (\frac{Q_+}{2t}\right ). \label{Eq:FFunc}
\end{equation}
In arriving at this result, we will, in the intermediate steps, also find a term proportional to $\tau_3\sigma_31$.  However, we can use the periodicity of the integrand to show that this term will be zero at the end.  We can also see that this must happen due to the symmetries of our system.  Imagine that we tried calculating the expectation value of an observable, which would be represented by a matrix $\tau_i\sigma_j s_k$.  This expectation value will only be non-zero if the matrix is proportional to one of the matrices appearing in the above identity, since said expectation value involves a trace of the product of the Green's function with the associated matrix.  If a term proportional to $\tau_3\sigma_31$ were present, then this means that we can have a finite expectation value of the associated observable, which would, in this case, correspond to the gap opened by an anomalous quantum Hall order parameter.  This order parameter breaks time reversal symmetry.  However, we should not be able to develop a finite expectation value of this observable because our Hamiltonian is time reversal invariant.
\end{widetext}

\section{Determination of RG flow equations}
\begin{figure}[ht]
\centering
\includegraphics[width=\columnwidth]{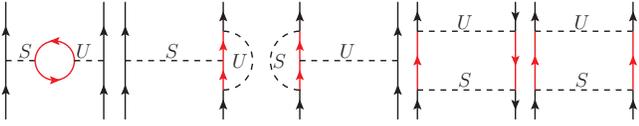}
\caption{\label{FFVertex_Renorm}Diagrams representing contributions to the renormalization of the four-fermion coupling constants $g_i$.  The dashed lines represent $8\times 8$ matrices, the black lines represent slow modes, and the red lines represent fast modes.}
\end{figure}
\begin{figure*}[t]
\centering
\includegraphics[width=0.75\textwidth]{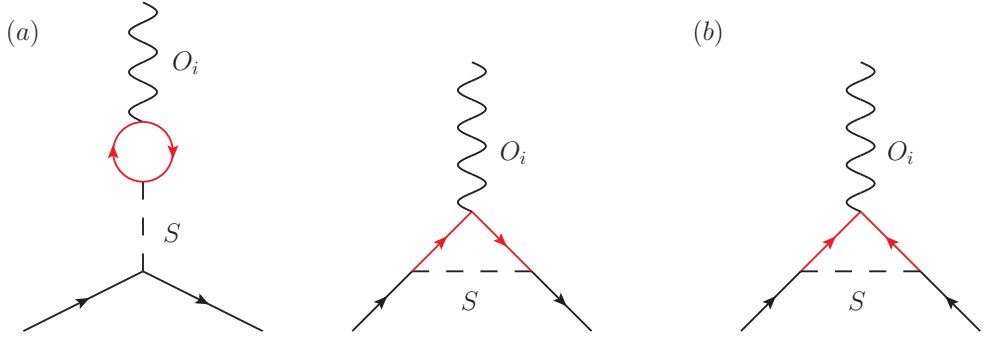}
\caption{\label{SourceVertex_Renorm}(a) Diagrams representing contributions to the renormalization of the particle-hole source terms.  All lines are as in Figure \ref{FFVertex_Renorm}.  In addition, the wavy lines represent the source terms.  (b) Diagram representing contributions to the renormalization of the particle-particle source terms.}
\end{figure*}
We now show how to derive the flow equations for the four-fermion coupling constants $g$ and the source term constants $\Delta$.  We start by performing a cumulant expansion of the partition function to second order in the ``perturbation'' $S_{int}+\Delta S$.
\begin{equation}
Z\approx \exp\left [-\langle S_{int}+\Delta S\rangle_0+\tfrac{1}{2}\langle (S_{int}+\Delta S)^2\rangle_{0,C}\right ],
\end{equation}
where, in the subscripts on the averages $\langle\cdot\rangle$, ``$0$'' means to average with respect to the bare action $S_0$, and ``$C$'' means that the average is ``connected''; that is, it can be represented with connected Feynman diagrams.  We now integrate out modes in thin shells; by doing so, we generate terms that renormalize different constants in our theory.  We will first discuss the terms that renormalize the four-fermion coupling constants, since the general procedure is the same.  There are five different types of terms that appear; these are represented by the diagrams shown in Figure \ref{FFVertex_Renorm}.

\begin{widetext}
The first diagram gives the following correction:
\begin{equation}
\delta S_1=\tfrac{1}{2}\sum_{S,U}g_S g_U \int_{1,2,3,4} \left\{\int_{\bk_>,\omega}\mbox{Tr}[S1G_{\bk}(i\omega)U1G_{\bk}(i\omega)]\right\}\sum_{\sigma,\sigma'}\psi^{\dag}_\sigma(1)S\psi_\sigma(2)\psi^{\dag}_{\sigma'}(3)U\psi_{\sigma'}(4),
\end{equation}
where the numbers $1-4$ are shorthand for the momentum and frequency dependences of the Grassman fields, and $\int_{1,2,3,4}$ represents the integrals and sums over these variables along with the proper momentum- and frequency-conserving $\delta$ functions, and similarly for $\int_{\bk_>,\omega}$.  We may evaluate the integral and sum over $\bk$ and $\omega$, respectively, using Eq.\ \eqref{2GF_Identity} in the absence of an external electric field or Eq.\ \eqref{2GF_EField_Identity} when said field is present.  In both cases, this term is only nonzero if $S=U$, so that we generate a correction to $g_S$ that is proportional to $g_S^2$.  The nonzero contributions to the coefficients,
\begin{equation}
A_{ijk}^{(a)}=A_{ijk}^{(a)}(1)+A_{ijk}^{(a)}(2+3)+A_{ijk}^{(a)}(4)+A_{ijk}^{(a)}(5),
\end{equation}
in Eq.\ \eqref{eq:gflow} are
\begin{eqnarray}
A_{iii}^{(1/2)}(1)&=&-\tfrac{1}{2}\{8\pm\mbox{Tr}[(\Gamma_i^{(1)}\tau_3 1_4)^2]\}\frac{m^*}{4\pi}, \\
A_{iii}^{(3/4)}(1)&=&\tfrac{1}{4}\{\mbox{Tr}[(\Gamma_i^{(1)}1\sigma_11)^2]\mp\mbox{Tr}[(\Gamma_i^{(1)}\tau_3\sigma_11)^2]\mp\mbox{Tr}[(\Gamma_i^{(1)}1\sigma_21)^2]+\mbox{Tr}[(\Gamma_i^{(1)}\tau_3\sigma_21)^2]\}\frac{m^*}{4\pi}, \\
A_{iii}^{(5/6)}(1)&=&\tfrac{1}{2}\{\mbox{Tr}[(\Gamma_i^{(1)}1\sigma_31)^2]\pm\mbox{Tr}[(\Gamma_i^{(1)}\tau_3\sigma_31)^2]\}\frac{m^*}{4\pi}.
\end{eqnarray}
In these expressions, the top signs correspond to the first number in the superscript on the left-hand side, while the bottom corresponds to the second.  The $A_{ijk}^{(5/6)}$ coefficients only enter into our analysis when a finite electric field is present.  The $8\times 8$ matrices $\Gamma_i^{(m)}$ are defined as follows:
\begin{eqnarray}
\Gamma_1^{(1)}&=&1_8 \label{Eq:GammaMat1} \\
\Gamma_2^{(1)}&=&\tau_3\sigma_31 \\
\Gamma_3^{(1)}&=&1\sigma_11,\,\Gamma_3^{(2)}=\tau_3\sigma_21 \\
\Gamma_4^{(1)}&=&\tau_31_4 \\
\Gamma_5^{(1)}&=&1\sigma_31 \\
\Gamma_6^{(1)}&=&\tau_3\sigma_11,\,\Gamma_6^{(2)}=-1\sigma_21 \\
\Gamma_7^{(1)}&=&\tau_1\sigma_11,\,\Gamma_7^{(2)}=\tau_2\sigma_11 \\
\Gamma_8^{(1)}&=&\tau_1\sigma_21,\,\Gamma_8^{(2)}=\tau_2\sigma_21 \\
\Gamma_9^{(1)}&=&\tau_11_4,\,\Gamma_9^{(2)}=-\tau_2\sigma_31,\,\Gamma_9^{(3)}=-\tau_21_4,\,\Gamma_9^{(4)}=-\tau_1\sigma_31. \label{Eq:GammaMat9}
\end{eqnarray}
The superscripts $(m)$ refer to the multiplicity of a given representation.  Here, and throughout this appendix, $A_{ijk}^{(a)}(n)$ represents the contribution to $A_{ijk}^{(a)}$ from the $n$th diagram in Fig.\ \ref{FFVertex_Renorm}.

The second and third diagrams together give the following correction:
\begin{equation}
\delta S_{2+3}=-\sum_{S,U}g_S g_U \int_{1,2,3,4} \psi^{\dag}_\sigma(1)S\psi_\sigma(2)\psi^{\dag}(3)\left [\int_{\bk_>,\omega}U1G_{\bk}(i\omega)S1G_{\bk}(i\omega)U1\right ]\psi(4).
\end{equation}
Note that the first two $\psi$ fields carry an explicit spin index.  The second two do not; for notational simplicity, these two are extended to be eight-component spinors in valley, layer, and spin space.  In both cases that we consider, the second matrix $U1G_{\bk}(i\omega)S1G_{\bk}(i\omega)U1$ appearing in this expression is proportional to $S1$.  Therefore, this term also represents a correction to $g_S$, but now it generates terms involving the products, $g_S g_U$.  We may extract the contributions to the $A_{ijk}^{(a)}$ coefficients by noting that, since the second matrix is proportional to $S1$.  Using $\mbox{Tr}(\Gamma_i^{(m)}\Gamma_j^{(n)})=8\delta_{ij}\delta_{mn}$, we find that the nonzero contributions to the $A_{ijk}^{(a)}$ coefficients are
\begin{eqnarray}
A_{iij}^{(1/2)}(2+3)&=&\tfrac{1}{8}\sum_{m=1}^{m_j}\{\mbox{Tr}[(\Gamma_i^{(1)}\Gamma_j^{(m)})^2]\pm\mbox{Tr}(\Gamma_i^{(1)}\Gamma_j^{(m)}\tau_3 1_4\Gamma_i^{(1)}\tau_3 1_4\Gamma_j^{(m)})\}\frac{m^*}{4\pi}, \\
A_{iij}^{(3/4)}(2+3)&=&-\tfrac{1}{16}\sum_{m=1}^{m_j}[\mbox{Tr}(\Gamma_i^{(1)}\Gamma_j^{(m)}1\sigma_11\Gamma_i^{(1)}1\sigma_11\Gamma_j^{(m)})\mp\mbox{Tr}(\Gamma_i^{(1)}\Gamma_j^{(m)}\tau_3\sigma_11\Gamma_i^{(1)}\tau_3\sigma_11\Gamma_j^{(m)}) \cr
&\mp&\mbox{Tr}(\Gamma_i^{(1)}\Gamma_j^{(m)}1\sigma_21\Gamma_i^{(1)}1\sigma_21\Gamma_j^{(m)})+\mbox{Tr}(\Gamma_i^{(1)}\Gamma_j^{(m)}\tau_3\sigma_21\Gamma_i^{(1)}\tau_3\sigma_21\Gamma_j^{(m)})]\frac{m^*}{4\pi}, \\
A_{iij}^{(5/6)}(2+3)&=&-\tfrac{1}{8}\sum_{m=1}^{m_j}[\mbox{Tr}(\Gamma_i^{(1)}\Gamma_j^{(m)}1\sigma_31\Gamma_i^{(1)}1\sigma_31\Gamma_j^{(m)})\pm\mbox{Tr}(\Gamma_i^{(1)}\Gamma_j^{(m)}\tau_3\sigma_31\Gamma_i^{(1)}\tau_3\sigma_31\Gamma_j^{(m)})]\frac{m^*}{4\pi}.
\end{eqnarray}
Here, the sum on $m$ is taken over the multiplicity of the $j$th representation, and the ``$2+3$'' in the ``arguments'' means that the given contribution is the total contribution from the second and third diagrams.

Finally, the fourth and fifth diagrams give the following:
\begin{eqnarray}
\delta S_4&=&-\tfrac{1}{2}\sum_{S,U}g_S g_U \int_{1,2,3,4} \int_{\bk_>,\omega}\psi^{\dag}(1)SG_{\bk}(i\omega)U\psi(2)\psi^{\dag}(3)UG_{\bk}(i\omega)S\psi(4) \\
\delta S_5&=&-\tfrac{1}{2}\sum_{S,U}g_S g_U \int_{1,2,3,4} \int_{\bk_>,\omega}\psi^{\dag}(1)SG_{\bk}(i\omega)U\psi(2)\psi^{\dag}(3)SG_{-\bk}(-i\omega)U\psi(4)
\end{eqnarray}
Both matrices occurring in each expression are proportional to each other, but will in general not be proportional to either $S$ or $U$.  These terms therefore represent corrections to a coupling $g_V$ that are proportional to $g_S g_U$.  Using the same observation as before, we can find the contributions to the $A_{ijk}^{(a)}$ coefficients.  Denoting $V=\Gamma_k$, these are
\begin{eqnarray}
A_{kij}^{(1/2)}(4)&=&\tfrac{1}{128}\sum_{m=1}^{m_i}\sum_{n=1}^{m_j}[\mbox{Tr}(\Gamma_k^{(1)}\Gamma_i^{(m)}\Gamma_j^{(n)})\mbox{Tr}(\Gamma_k^{(1)}\Gamma_j^{(n)}\Gamma_i^{(m)})\pm\mbox{Tr}(\Gamma_k^{(1)}\Gamma_i^{(m)}\tau_3 1_4\Gamma_j^{(n)})\mbox{Tr}(\Gamma_k^{(1)}\Gamma_j^{(n)}\tau_3 1_4\Gamma_i^{(m)})]\frac{m^*}{4\pi}, \\
A_{kij}^{(3/4)}(4)&=&-\tfrac{1}{256}\sum_{m=1}^{m_i}\sum_{n=1}^{m_j}[\mbox{Tr}(\Gamma_k^{(1)}\Gamma_i^{(m)}1\sigma_11\Gamma_j^{(n)})\mbox{Tr}(\Gamma_k^{(1)}\Gamma_j^{(n)}1\sigma_11\Gamma_i^{(m)})\mp\mbox{Tr}(\Gamma_k^{(1)}\Gamma_i^{(m)}\tau_3\sigma_11\Gamma_j^{(n)})\mbox{Tr}(\Gamma_k^{(1)}\Gamma_j^{(n)}\tau_3\sigma_11\Gamma_i^{(m)}) \cr
&\mp&\mbox{Tr}(\Gamma_k^{(1)}\Gamma_i^{(m)}1\sigma_21\Gamma_j^{(n)})\mbox{Tr}(\Gamma_k^{(1)}\Gamma_j^{(n)}1\sigma_21\Gamma_i^{(m)})+\mbox{Tr}(\Gamma_k^{(1)}\Gamma_i^{(m)}\tau_3\sigma_21\Gamma_j^{(n)})\mbox{Tr}(\Gamma_k^{(1)}\Gamma_j^{(n)}\tau_3\sigma_21\Gamma_i^{(m)})]\frac{m^*}{4\pi}, \\
A_{kij}^{(5/6)}(4)&=&\tfrac{-1}{128}\sum_{m=1}^{m_i}\sum_{n=1}^{m_j}[\mbox{Tr}(\Gamma_k^{(1)}\Gamma_i^{(m)}1\sigma_31\Gamma_j^{(n)})\mbox{Tr}(\Gamma_k^{(1)}\Gamma_j^{(n)}1\sigma_31\Gamma_i^{(m)})\pm\mbox{Tr}(\Gamma_k^{(1)}\Gamma_i^{(m)}\tau_3\sigma_31\Gamma_j^{(n)})\mbox{Tr}(\Gamma_k^{(1)}\Gamma_j^{(n)}\tau_3\sigma_31\Gamma_i^{(m)})]\frac{m^*}{4\pi}, \nonumber \\
\end{eqnarray}
and
\begin{eqnarray}
A_{kij}^{(1/2)}(5)&=&-\tfrac{1}{128}\sum_{m=1}^{m_i}\sum_{n=1}^{m_j}\{[\mbox{Tr}(\Gamma_k^{(1)}\Gamma_i^{(m)}\Gamma_j^{(n)})]^2\mp[\mbox{Tr}(\Gamma_k^{(1)}\Gamma_i^{(m)}\tau_3 1_4\Gamma_j^{(n)})]^2\}\frac{m^*}{4\pi}, \\
A_{kij}^{(3/4)}(5)&=&-\tfrac{1}{256}\sum_{m=1}^{m_i}\sum_{n=1}^{m_j}\{[\mbox{Tr}(\Gamma_k^{(1)}\Gamma_i^{(m)}1\sigma_11\Gamma_j^{(n)})]^2\pm[\mbox{Tr}(\Gamma_k^{(1)}\Gamma_i^{(m)}\tau_3\sigma_11\Gamma_j^{(n)})]^2 \cr
&\pm&[\mbox{Tr}(\Gamma_k^{(1)}\Gamma_i^{(m)}1\sigma_21\Gamma_j^{(n)})]^2+[\mbox{Tr}(\Gamma_k^{(1)}\Gamma_i^{(m)}\tau_3\sigma_21\Gamma_j^{(n)})]^2\}\frac{m^*}{4\pi}, \\
A_{kij}^{(5/6)}(5)&=&-\tfrac{1}{128}\sum_{m=1}^{m_i}\sum_{n=1}^{m_j}\{[\mbox{Tr}(\Gamma_k^{(1)}\Gamma_i^{(m)}1\sigma_31\Gamma_j^{(n)})]^2\pm[\mbox{Tr}(\Gamma_k^{(1)}\Gamma_i^{(m)}\tau_3\sigma_31\Gamma_j^{(n)})]^2\}\frac{m^*}{4\pi}.
\end{eqnarray}

We now turn our attention to the symmetry-breaking source terms.  In this case, we have different procedures for the case without an applied electric field and the case with one.  We will consider the former case first.  The corrections to the particle-hole and particle-particle source terms are represented by the diagrams in Fig.\ \ref{SourceVertex_Renorm}.

The particle-hole source term corrections give us
\begin{eqnarray}
\delta S_{ph}&=&\sum_i \sum_S \Delta^{ph}_i g_S \int_{\bk'_<,\omega'}\int_{\bk_>,\omega}\mbox{Tr}[G_{\bk}(i\omega)O^{(i)}G_{\bk}(i\omega)S1]\psi^{\dag}_{\bk'}(\omega')S1\psi_{\bk'}(\omega') \cr
&-&\sum_i \sum_S \Delta^{ph}_i g_S \int_{\bk'_<,\omega'}\int_{\bk_>,\omega}\psi^{\dag}_{\bk'}(\omega')S1G_{\bk}(i\omega)O^{(i)}G_{\bk}(i\omega)S1\psi_{\bk'}(\omega').
\end{eqnarray}
In the first term, the trace will only be nonzero if $S1=O^{(i)}$, and, in the second term, the matrix appearing in the expression is proportional to $O^{(i)}$.  Therefore we see that different source terms are not mixed to this order.  Note that the first term is only nonzero if $O^{(i)}$ represents a charge order, and vanishes for spin orders.  The contributions to the coefficients
\begin{equation}
B^{(a)}_{ij}=B^{(a)}_{ij}(1)+B^{(a)}_{ij}(2) \label{Eq:BDef}
\end{equation}
in Equation \eqref{dlogDeltaph} are
\begin{eqnarray}
B^{(1/2)}_{ij}(1)&=&-\tfrac{1}{2}\sum_{n=1}^{m_j}[\mbox{Tr}(O^{(i)}\Gamma_j^{(n)})\pm\mbox{Tr}(\tau_3 1_4O^{(i)}\tau_3 1_4\Gamma_j^{(n)})]\frac{m^*}{4\pi}, \label{Eq:B121} \\
B^{(3/4)}_{ij}(1)&=&\tfrac{1}{4}\sum_{n=1}^{m_j}[\mbox{Tr}(1\sigma_11O^{(i)}1\sigma_11\Gamma_j^{(n)})\mp\mbox{Tr}(\tau_3\sigma_11O^{(i)}\tau_3\sigma_11\Gamma_j^{(n)}) \cr
&\mp&\mbox{Tr}(1\sigma_21O^{(i)}1\sigma_21\Gamma_j^{(n)})+\mbox{Tr}(\tau_3\sigma_21O^{(i)}\tau_3\sigma_21\Gamma_j^{(n)})]\frac{m^*}{4\pi}, \\
B^{(1/2)}_{ij}(2)&=&\tfrac{1}{16}\sum_{n=1}^{m_j}\{\mbox{Tr}[(O^{(i)}\Gamma_j^{(n)})^2]\pm\mbox{Tr}(O^{(i)}\Gamma_j^{(n)}\tau_3 1_4O^{(i)}\tau_3 1_4\Gamma_j^{(n)})\}\frac{m^*}{4\pi}, \label{Eq:B122} \\
B^{(3/4)}_{ij}(2)&=&-\tfrac{1}{32}\sum_{n=1}^{m_j}[\mbox{Tr}(O^{(i)}\Gamma_j^{(n)}1\sigma_11O^{(i)}1\sigma_11\Gamma_j^{(n)})\mp\mbox{Tr}(O^{(i)}\Gamma_j^{(n)}\tau_3\sigma_11O^{(i)}\tau_3\sigma_11\Gamma_j^{(n)}) \cr
&\mp&\mbox{Tr}(O^{(i)}\Gamma_j^{(n)}1\sigma_21O^{(i)}1\sigma_21\Gamma_j^{(n)})+\mbox{Tr}(O^{(i)}\Gamma_j^{(n)}\tau_3\sigma_21O^{(i)}\tau_3\sigma_21\Gamma_j^{(n)})]\frac{m^*}{4\pi}. \label{Eq:B342}
\end{eqnarray}
Here, the ``arguments'' have the same meaning as before, but with respect to Fig.\ \ref{SourceVertex_Renorm}.

The correction to the particle-particle source term is
\begin{eqnarray}
\delta S_{pp}=- \frac 1 2 \sum_{i=1}^{16} \sum_S \Delta^{pp}_i g_S \int_{\bk'_<,\omega'}\int_{\bk_>,\omega}\psi_{\bk'}^\dag(\omega')S1G_{\bk}(i\omega){\tilde O}^{(i)}[G_{-\bk}(-i\omega)]^T(S1)^T\psi_{-\bk'}^\ast(-\omega')+c.c.
\end{eqnarray}
For similar reasons as above, the product of five matrices appearing in this expression is proportional to ${\tilde O}^{(i)}$, and therefore different source terms are not mixed to this order.  Also note that the $8\times 8$ matrix ${\tilde O}^{(i)}$ must be completely antisymmetric.  The values of the coefficients ${\tilde B}^{(a)}_{ij}$ in Equation \eqref{dlogDeltapp} are therefore
\begin{eqnarray}
{\tilde B}^{(1/2)}_{ij}&=&-\tfrac{1}{16}\sum_{n=1}^{m_j}\{\mbox{Tr}[{\tilde O}^{(i)}\Gamma_j^{(n)}{\tilde O}^{(i)}(\Gamma_j^{(n)})^T]\mp\mbox{Tr}[{\tilde O}^{(i)}\Gamma_j^{(n)}\tau_3 1_4{\tilde O}^{(i)}\tau_3 1_4(\Gamma_j^{(n)})^T]\}\frac{m^*}{4\pi}, \label{Eq:Bt12} \\
{\tilde B}^{(3/4)}_{ij}&=&-\tfrac{1}{32}\sum_{n=1}^{m_j}\{\mbox{Tr}[{\tilde O}^{(i)}\Gamma_j^{(n)}1\sigma_11{\tilde O}^{(i)}1\sigma_11(\Gamma_j^{(n)})^T]\pm\mbox{Tr}[{\tilde O}^{(i)}\Gamma_j^{(n)}\tau_3\sigma_11{\tilde O}^{(i)}\tau_3\sigma_11(\Gamma_j^{(n)})^T] \cr
&\mp&\mbox{Tr}[{\tilde O}^{(i)}\Gamma_j^{(n)}1\sigma_21{\tilde O}^{(i)}1\sigma_21(\Gamma_j^{(n)})^T]-\mbox{Tr}[{\tilde O}^{(i)}\Gamma_j^{(n)}\tau_3\sigma_21{\tilde O}^{(i)}\tau_3\sigma_21(\Gamma_j^{(n)})^T]\}\frac{m^*}{4\pi}. \label{Eq:Bt34}
\end{eqnarray}
\end{widetext}

\begin{figure}[ht]
\centering
\includegraphics[width=\columnwidth]{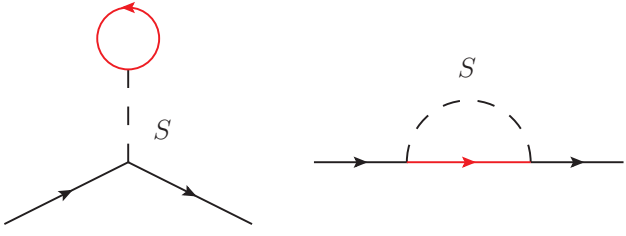}
\caption{\label{EFVertex_Renorm}Diagrams representing contributions to the renormalization of the applied electric field term.  All lines are as in Figure \ref{SourceVertex_Renorm}.}
\end{figure}

Now we consider corrections to the {\em finite} applied electric field; see Eq. \eqref{Eq:VFlow}.  In this case, we find that the lowest-order corrections come from the {\em first-order} term in the cumulant expansion; these first-order corrections would be zero in the absence of the electric field.  They are represented by ``tadpole'' and ``sunrise'' diagrams, as shown in Figure \ref{EFVertex_Renorm}.
The contribution from the ``tadpole'' diagrams is
\begin{equation}
\delta S_t=-\sum_S g_S\int_{\bk'_{<},\omega'} \int_{\bk_{>},\omega}\!\!\!\mbox{Tr}[S1G_{\bk}(i\omega)]\psi^\dag_{\bk'}(\omega')S1\psi_{\bk'}(\omega').
\end{equation}
The integral over $\bk$ and sum over $\omega$ can be evaluated using Equation \eqref{1GF_Identity}.  The trace occurring in this expression is only nonzero if $S=1\sigma_3$.  Therefore, we only generate a correction to the applied electric field.  Since $1\sigma_31=\Gamma_5^{(1)}$, we see that the only non-zero contribution from this term to the coefficients $b_i$ in Eq. \eqref{Eq:VFlow} is to $b_5$, and this contribution is $b_5(\text{tadpole})=8\times\frac{m^*}{4\pi}$.

The ``sunrise'' diagrams give us
\begin{equation}
\delta S_s=\sum_S g_S\int_{\bk'_{<},\omega'} \int_{\bk_{>},\omega}\psi^\dag_{\bk'}(\omega')S1G_{\bk}(i\omega)S1\psi_{\bk'}(\omega').
\end{equation}
The matrix occurring in this expression is proportional to $1\sigma_31$, and thus we, once again, only generate corrections to the applied electric field.  This will contribute to all of the $b_i$.  These contributions are given by
\begin{equation}
b_i(\text{sunrise})=\tfrac{1}{8}\sum_{m}\mbox{Tr}(1\sigma_31\Gamma_i^{(m)} 1\sigma_31\Gamma_i^{(m)})\frac{m^*}{4\pi}.
\end{equation}
The total value of $b_i$ is simply the sum of the above two contributions, i.e., $b_i=b_i(\text{tadpole})+b_i(\text{sunrise})$.

\section{Coefficients in the free energy} \label{App:Coeff_FE}
The coefficients $\alpha_{a,i}^{ph}$ appearing in the free energy, Eq. \eqref{eq:free energy}, are
\begin{eqnarray}
\alpha_{1/2,i}^{ph}&=&8\pm\mbox{Tr}[(O^{(i)}\tau_{3}1_4)^2], \label{Eq:alpha12ph} \\
\alpha_{3/4,i}^{ph}&=&-\tfrac{1}{2}\{\mbox{Tr}[(O^{(i)}1\sigma_11)^2]\mp\mbox{Tr}[(O^{(i)}\tau_3\sigma_11)^2] \cr
&\mp&\mbox{Tr}[(O^{(i)}1\sigma_21)^2]+\mbox{Tr}[(O^{(i)}\tau_3\sigma_21)^2]\}.
\end{eqnarray}
The $\alpha_{a,i}^{pp}$ coefficients are
\begin{eqnarray}
\alpha_{1/2,i}^{pp}&=&8\mp\mbox{Tr}[({\tilde O}^{(i)}\tau_{3}1_4)^2], \\
\alpha_{3/4,i}^{pp}&=&\tfrac{1}{2}\{\mbox{Tr}[({\tilde O}^{(i)}1\sigma_11)^2]\pm\mbox{Tr}[({\tilde O}^{(i)}\tau_3\sigma_11)^2] \cr
&\mp&\mbox{Tr}[({\tilde O}^{(i)}1\sigma_21)^2]-\mbox{Tr}[({\tilde O}^{(i)}\tau_3\sigma_21)^2]\}. \label{Eq:alpha34pp}
\end{eqnarray}

\section{Analytic determination of the phase boundaries in the fixed ratio plane}
We will now describe how we determined the phase boundaries in the
target plane.  These boundaries are defined by the sign of the
susceptibility exponent $\gamma_i$, as given by Equation
\eqref{SuscExp_Formula}, for a given phase; whenever it is positive,
we say that the associated phase is present.  The value of
$\mathcal{A}_{(E_g)}$ is given by Equation \eqref{FixedPlane_A}. We may obtain
$\mathcal{B}_{i,(E_g)}$ from Eqs. \eqref{eq:BCoeffPH} and \eqref{eq:BCoeffPP}
and from the coupling constant ratios $\rho_i^{(E_g)}$ given in Equations
\eqref{FixedPlane_gA1g}-\eqref{FixedPlane_gA2K}.  Because of this,
all of the $\gamma_i$ will have the form,
\begin{equation}
\gamma_i=\frac{Q_i(x,y)}{3+2x+3x^2+4y+4xy+8y^2}, \label {gammaConditions}
\end{equation}
where $Q_i(x,y)$ is an inhomogeneous quadratic function of $x$ and
$y$.  The denominator of this expression is positive definite, so
that the sign of the exponent is determined entirely by $Q_i(x,y)$.
Our condition that $\gamma_i$ be positive thus requires that
$Q_i(x,y)>0$.  We therefore see that the phase boundaries, given by
$Q_i(x,y)=0$, are all conic sections.


\begin{thebibliography}{99}
\bibitem{MWargument} Due to the Mermin-Wagner argument, if there is a coexistence of nematic with AF/QSH, then it can only appear at $T=0$, since, unlike the nematic, which breaks a discrete symmetry ($D_{3d}\rightarrow C_i$), AF and QSH break a continuous $O(3)$ symmetry. We see no reason why such a putative line could not end at a finite value of $g_{A_{1g}}$ marked by the question mark.
\end{thebibliography}
\end{document}